\documentclass[superscriptaddress,
nofootinbib,
nobibnotes,
notitlepage,
amsmath,amssymb,
aps,
twocolumn,
prd,
]{revtex4-2}
\usepackage[T1]{fontenc}
\usepackage[utf8]{inputenc}
\usepackage{bm}
\usepackage{color}
\usepackage{xspace}
\usepackage{multirow}
\usepackage{rotating}
\usepackage{threeparttable}
\usepackage{tabularx}
\usepackage{lineno}
\usepackage{fontawesome5}
\usepackage{hyperref}

\def\beq{\begin{eqnarray}}
\def\eeq{\end{eqnarray}}

\usepackage[normalem]{ulem}
\usepackage{physics}
\usepackage{ifthen}
\usepackage{aas_macros}
\usepackage{dcolumn}
\usepackage{orcidlink}
\newcolumntype{d}{D{.}{.}{-1}}
\newcolumntype{C}{>{$}c<{$}}
\graphicspath{{}}

\usepackage{siunitx}

\begin{document}

\preprint{APS/123-QED}

\newcommand{\Hunit}{{\rm km}\ {\rm s}^{-1}\rm{Mpc}^{-1}}
\newcommand{\Frank}[1]{\textcolor{blue}{(FJQ: #1)}}
\newcommand{\hun}{\,\mathrm{km}\,\mathrm{s}^{-1}\mathrm{Mpc}^{-1}}
\newcommand{\es}[1]{\textcolor{blue}{(ES: #1)}}
\newcommand{\kw}[1]{\textcolor{purple}{#1}}
\newcommand{\KW}[1]{\textcolor{purple}{(KW: #1)}}
\newcommand{\hanyue}[1]{\textcolor{pink}{(HW: #1)}}
\newcommand{\fge}[1]{\textcolor{blue}{#1}}
\newcommand{\fei}[1]{\textcolor{teal}{#1}}
\newcommand{\mgeo}{\sum m_\nu^{\rm geo}}
\newcommand{\mgrowth}{\sum m_\nu^{\rm growth}}
\newcommand{\meff}{\sum m_\nu^{\rm eff}}
\newcommand{\mdiff}{\Delta\sum m_\nu}
\newcommand{\mnu}{\sum m_\nu}


\title{
Disentangling geometry \& growth in cosmic neutrino mass constraints: 
a more physical split and diagnostic
}

\author{Fei Ge\,\orcidlink{0000-0002-3833-8133}}
\email{feige@caltech.edu}
\affiliation{Department of Physics, California Institute of Technology, Pasadena, CA, 91125, USA}
\affiliation{Kavli Institute for Particle Astrophysics and Cosmology, Stanford University, 452 Lomita Mall, Stanford, CA, 94305, USA}
\affiliation{Department of Physics, Stanford University, 382 Via Pueblo Mall, Stanford, CA, 94305, USA}

\author{Frank J. Qu \,\orcidlink{0000-0001-7805-1068}}
\affiliation{Kavli Institute for Particle Astrophysics and Cosmology, Stanford University, 452 Lomita Mall, Stanford, CA, 94305, USA}
\affiliation{Department of Physics, Stanford University, 382 Via Pueblo Mall, Stanford, CA, 94305, USA}

\author{Hanyue~Wang\,\orcidlink{0009-0009-3082-8021}}
\affiliation{Kavli Institute for Particle Astrophysics and Cosmology, Stanford University, 452 Lomita Mall, Stanford, CA, 94305, USA}
\affiliation{Department of Physics, Stanford University, 382 Via Pueblo Mall, Stanford, CA, 94305, USA}

\author{Emmanuel~Schaan\,\orcidlink{0000-0002-4619-8927}}
\affiliation{Kavli Institute for Particle Astrophysics and Cosmology, Stanford University, 452 Lomita Mall, Stanford, CA, 94305, USA}
\affiliation{SLAC National Accelerator Laboratory, 2575 Sand Hill Road, Menlo Park, California 94025, USA}

\author{W.~L.~Kimmy Wu\,\orcidlink{0000-0001-5411-6920}}
\affiliation{Department of Physics, California Institute of Technology, Pasadena, CA, 91125, USA}
\affiliation{Kavli Institute for Particle Astrophysics and Cosmology, Stanford University, 452 Lomita Mall, Stanford, CA, 94305, USA}
\affiliation{SLAC National Accelerator Laboratory, 2575 Sand Hill Road, Menlo Park, California 94025, USA}

\author{Alexander~Friedland\,\orcidlink{0000-0002-5047-4680}}
\affiliation{SLAC National Accelerator Laboratory, 2575 Sand Hill Road, Menlo Park, California 94025, USA}


\date{\today}

\begin{abstract}

Recent cosmological data hint at
a neutrino mass sum ($\mnu$) below the minimum value (0.058~eV) allowed by the neutrino mass oscillation experiments, with a preference for the opposite of massive neutrinos (``negative'' neutrino masses).
Since massive neutrinos impact the expansion history of the Universe, which in turn modifies the growth of structure, it is natural to ask whether such a preference originates from the observed geometry or structure growth,
and whether both effects appear consistent.
In this paper, for the first time, we introduce 
a more physical and rigorous split of $\mnu$ into parameters $\mgeo$ and $\mgrowth$ based on the corresponding physical effects, rather than on observables which often mix them.
We further extend such a split parameterization to negative values, as a diagnostic for geometry or growth preferring the opposite effect from massive neutrinos. 
Both channels are mediated by the expansion rate, so we split the expansion rate at the Boltzmann-code level: $\mgeo$ sets $H(z)$ wherever a distance is computed, and $\mgrowth$ sets $H(z)$ in the evolution of all the perturbations. 
This reduces to $\Lambda$CDM$+\sum m_\nu$ when $\mgeo = \mgrowth$.
Combining the {\it Planck} primary CMB, the {\it Planck}+ACT+SPT CMB lensing and the DESI BAO data, we find a similar sensitivity and ``negative'' preference from geometry and growth with
$\mgeo = -0.07\pm0.10$~eV (68\% C.L.) 
and 
$\mgrowth = -0.13^{+0.10}_{-0.12}$~eV (68\% C.L.). 
Since $\Lambda$CDM requires the two to be equal, their difference is a null test of the model, which is passed for current datasets: $\mdiff \equiv \mgeo - \mgrowth = 0.07\pm0.17$~eV (68\% C.L.), consistent with zero at $0.4\sigma$. 
For joint constraints with upcoming Advanced SO CMB data and DESI DR3 BAO data, forecast shows $\sigma(\mgeo)\sim0.04~\mathrm{eV}$ while $\sigma(\mgrowth)\sim0.08~\mathrm{eV}$. If the negative central value of $\mgeo = -0.07$~eV persists, such data combination would reject $0.058$~eV at $3.0\sigma$ from the geometry channel alone.
The two channels also respond differently to model error, thus providing a powerful new diagnostic between possible origins of the neutrino mass tension.
With future CMB and BAO data, our Fisher forecast show that an incorrect dark energy model displaces $\mgeo$ from $0.058$~eV by $3.1$--$4.2\sigma$ and $\mgrowth$ by $1.1$--$1.3\sigma$, while an underestimated reionization optical depth reverses the pattern, $2.7$--$3.6\sigma$ in $\mgrowth$ against at most $0.23\sigma$ in $\mgeo$. 
We make our modified Boltzmann code publicly available at \href{https://github.com/shuimufge/CAMB_geo_growth.git}{\faGithub~GitHub}.
\end{abstract}

\maketitle

\section{\label{sec:level1}Introduction}

Cosmological observations are uniquely positioned to measure the sum of neutrino masses. Although neutrino oscillation experiments provide evidence that neutrinos are massive, they only tightly constrain the mass-squared differences among neutrino flavors, from which the minimum neutrino mass sum of the inverted (IO) and normal (NO) mass ordering can be derived. On the other hand, cosmological observables are sensitive to the neutrino mass sum ($\mnu$) through its contribution to the total energy density, impacting both the expansion history and the structure growth of the Universe (see e.g.~\citep{Lesgourgues:2006nd, Lesgourgues_2012}).

The effects on cosmological observables are imprinted through two physically distinct channels. Under the current constraints, neutrinos become non-relativistic during the matter-dominated era and modify the expansion rate $H(z)$.
At the background level, this alters the angular diameter distance to the effective redshifts of cosmological observables, which is purely geometrical information. Since  structure growth mostly happens during matter domination, with the transition of massive neutrinos from relativistic to non-relativistic during this time, the mass of neutrinos also affects the large scale structure formation.

Cosmological observations probe both geometry and growth of the Universe. Baryon acoustic oscillations (BAO) is a robust geometrical probe by measuring the angular scale of the sound waves of the photon-baryon plasma, enabling precise measurement of the cosmic expansion \citep{Blake:2003rh, Seo:2003pu, weinberg2013observational}. Cosmic microwave background (CMB) precisely encodes the anisotropic patterns of the photons emitted from the last scattering surface when photon decouples from the baryons, containing both geometrical and growth information in its primary anisotropy and the gravitational lensing of the CMB photons \citep{Hu:2001bc, Hu:1995en, Efstathiou:1998xx,Lewis:2006fu}.

Precise cosmological surveys of BAO and CMB can provide tight constraints on $\mnu$. Recently, several cosmological observations probing $\mnu$ have been released. These include probes of CMB anisotropy from {\it Planck}, the Atacama Cosmology Telescope (ACT) \citep{ACT:2025xdm,ACT:2025tim,ACT:2025fju}, the South Pole Telescope \citep{camphuis2025spt3gd1cmbtemperature}, and late-universe probes from galaxy surveys such as the Dark Energy Spectroscopic Instrument (DESI) \citep{garciaquintero2025cosmologicalimplicationsdesidr2, desicollaboration2025desidr2resultsii} 
and CMB lensing from {\it Planck}, ACT, SPT \citep{qu2025unifiedconsistentstructuregrowth}. The joint analysis shows $\sum m_\nu < 0.055 \ \mathrm{eV \ (95\% C.L.)}$\citep{camphuis2025spt3gd1cmbtemperature}. It hints a surprising anomaly where the inferred cosmological neutrino mass is lower than the minimum value allowed by neutrino oscillation experiments at the $95\% \ \mathrm{C.L.}$ level\footnote{The latest SPT-D1 CMB lensing likelihood, combined with primary CMB and BAO, infers $\sum m_\nu < 0.072 \ \mathrm{eV \ (95\% C.L.)}$.}.

This unphysical result raises questions about potential data systematics, new physics interpretations, subtle model dependencies, and the need for improved methods to interpret and combine datasets. 
Many results have emerged attempting to explain this apparent tension. In \citep{Loverde:2024nfi,lynch2025whatsmattersigmamnu}, the authors argued that the leading cause of this neutrino mass upper limit tension is due to the mismatch between the total matter density preferred by BAO with the cold dark matter (CDM) and baryon density inferred from the CMB observables. On the other hand, CMB lensing measurement was found to drive the preference for the anomalous $\mnu$ in \citep{craig2024nusgoodnews,PhysRevD.111.083507} when $\mnu$ is used to only scale the CMB lensing power spectrum. In \cite{Sailer:2025lxj,Jhaveri:2025neg,Sullivan:2026tas}, a higher value of reionization optical depth is found to be able to relieve the $\mnu$ tension inferred from CMB+BAO with the minimum value from neutrino oscillation experiments. 

In this work, we attempt to identify, between the cosmological information from the geometry and growth channel, which channel drives the constraint of $\sum m_\nu$ and tight upper bound of $\sum m_\nu$. Two possible causes to such a tight constraint are: 
(1) a genuine improvement of the cosmological observable measurement, 
or (2) a shift of the marginal posterior toward the unphysical negative regime, which truncates against the physical prior at zero and artificially narrows the visible part of the distribution. For parameters with Gaussian posterior distributions, these two effects can be separable; for the posterior distribution of $\sum m_\nu$, however, the two effects are blended together due to the truncation of the prior at zero, making it hard to identify the leading cause of the tightening of its posterior distribution.
Both the geometrical and perturbative information from the cosmological observables contribute to the $\mnu$ constraints, entangled with the two statistical effects above. 

To disentangle these two channels, we introduce two phenomenological parameters $\mgeo$ and $\mgrowth$ that quantify the geometrical and perturbative effects of the massive neutrinos on CMB and BAO observables. 
Our approach differs from prior work in three respects. First, unlike the data-based split of \citep{graham2025darkforcesgathering} into $\sum m_\nu^{\rm clustering}$ and $\sum m_\nu^{\rm BAO}$, we split by physical effect rather than by dataset, so both of our parameters appear in the CMB prediction and can be constrained jointly. Second, unlike the `background-only' likelihoods of \citep{Loverde:2024nfi, lynch2025whatsmattersigmamnu}, which marginalize over perturbative information at the likelihood level, our split is implemented at the Boltzmann-code level and cleanly isolates the two channels. Third, in spirit our parametrization resembles the geometry-growth consistency tests of \citep{Andrade:2021njl, Wang:2007fsa, Ruiz:2014hma, Bernal:2015zom, DES:2020iqt, Ruiz-Zapatero:2021rzl}, but applied to $\mnu$ rather than to dark energy or modified gravity.

Throughout the paper, we have fixed the effective number of neutrino species, $N_{\rm eff}$, to 3.044. The paper is organized as follows: 
We start with the baseline constraints on $\mnu$ and check if the neutrino mass sum anomaly can be attributed to a single cosmological observable. The CMB and BAO datasets are introduced in Sect.~\ref{sec:data}, followed by discussion of swapping the data vectors from observations with mock data vectors in Sect.~\ref{sec:fiducial_swap}. We briefly review the discussions of the causes to the $\sum m_\nu$ anomaly in Sect.~\ref{sec:review}, before we discuss the implementation of $\mgeo$ and $\mgrowth$ and their impact on cosmological observables in Sect.~\ref{sec:split_imp}. We discuss the results using the $\mgeo$ and $\mgrowth$ splits in Sect.~\ref{sec:split_results}, followed by forecasts of ongoing and future CMB and galaxy surveys in Sect.~\ref{sec:forecast}. We conclude in Sect.~\ref{sec:conclusion}.

\section{Tensions between CMB and BAO, and the need to split geometry VS growth}

In this section, we investigate if the preference for a value of $\mnu$ less than the lower limit from neutrino oscillation experiments is caused by a specific data set. Our results show that the tight current upper bound on $\sum m_\nu$ does not seem to originate from any single anomalous dataset. 
From this, we conclude that splitting the $\sum m_\nu$ based on datasets may not be the most promising approach, and that splitting based on the physical effects, geometry VS growth as in Sec.~\ref{sec:split_imp}, may prove insightful.

\subsection{Data}
\label{sec:data}

Before showing the results, we specify the data likelihoods used in this work. There are three cosmic probes used in this analysis:

\textbf{CMB Primary Spectrum}: We use the \texttt{CamSpec} likelihood of the \textit{Planck} NPIPE PR4 map for the $\ell>30$ section of the CMB likelihood \citep{Rosenberg:2022sdy}.
As a default choice, we also include the \texttt{Plik-lowl-TT} \citep{Planck:2019nip} and the \texttt{SRoll2} low-$\ell$ EE likelihood \citep{Delouis:2019bub}. We only drop these two low-$\ell$ likelihoods and use a Gaussian prior on the reionization optical depth, $\tau$, with dispersion of $0.006$ and mean at the best-fit values in Table~\ref{tab:mock_params} for the \texttt{mock} \texttt{CamSpec} runs in Sect.~\ref{sec:fiducial_swap}. We use \texttt{CMB} for the primary CMB spectra datasets. 

\textbf{CMB Lensing:} We use a Gaussian likelihood \citep{qu2025unifiedconsistentstructuregrowth}
 that combines the CMB lensing bandpowers from three experiments: ACT DR6, Planck PR4, and SPT-3G M2PM. 
 The combined likelihood yields the most precise CMB lensing power spectrum to date, with a joint signal-to-noise ratio of 61. The likelihood accounts for both auto-covariances within each experiment and cross-covariances between overlapping survey regions, with the covariance matrix estimated from simulations and analytic calculations. We label it with \texttt{Lensing} for short.

\textbf{BAO:} We use the BAO measurement from \texttt{DESI} DR2 \citep{desicollaboration2025desidr2resultsii} with the effective redshift range from 0.295 to 2.33 using all BGS, LRG, ELG, QSO and Lya samples. We use \texttt{BAO} for this data for short.

Of each probe, we use both the \texttt{default} likelihood and a \texttt{mock} likelihood. Compared to the \texttt{default} likelihood, we only replace the data vector in \texttt{mock} likelihood with the the predicted values from a mock $\Lambda$CDM cosmology, which is either the \texttt{CMB}-only best-fit or the \texttt{CMB+BAO} joint best-fit cosmology. The $\Lambda$CDM parameters used for generating the mock data are summarized in Table~\ref{tab:mock_params}. 

\begin{table}
\centering
\begin{tabular}{l | c  c}
\hline
Param. & $\mathtt{mock}_\mathrm{P}$ & $\mathtt{mock}_\mathrm{P+D}$ \\
\hline
$\theta_\mathrm{MC}$ & $0.01041$ & $0.01041$ \\
$\omega_\mathrm{b}$ & $0.02236$ & $0.02233$ \\
$\omega_\mathrm{c}$ & $0.1200$ & $0.1176$ \\
$\log(10^{10} A_\mathrm{s})$ & $3.051$ & $3.062$ \\
$n_s$ & $0.965$ & $0.969$ \\
$\tau$ & $0.057$ & $0.065$ \\
\hline
\end{tabular}
\caption{Cosmological parameters used for generating mock data. Two sets of mock data are generated: one with the best-fit parameters from \texttt{CMB}, labelled as $\mathtt{mock}_\mathrm{P}$, and another one with \texttt{CMB}+\texttt{BAO} best fit, labelled as $\mathtt{mock}_\mathrm{P+D}$.}
\label{tab:mock_params}
\end{table}

\subsection{Shifts in $\sum m_\nu$ Posterior Caused by Each Data}\label{sec:fiducial_swap}

In this section, we show that the mild disagreement among \texttt{CMB}, \texttt{BAO} and \texttt{Lensing} data sets causes the shift of the neutrino mass sum posterior toward negative values. 
We establish this by comparing the $\sum m_\nu$ posterior distributions using the \texttt{default} and \texttt{mock} likelihoods. Specifically, we replace individual data vectors with predictions from two different mock $\Lambda$CDM cosmologies: a \texttt{CMB}-only best fit, which assumes the best-fit \texttt{CMB} cosmology is the truth, and a joint \texttt{CMB}+\texttt{BAO} best fit, which does not privilege either dataset. The runs with \texttt{CMB}-only $\Lambda$CDM best-fit mock data provide tests of whether \texttt{BAO} and \texttt{lensing} are individually consistent with \texttt{CMB}. The second one using the joint \texttt{CMB}+\texttt{BAO} $\Lambda$CDM best-fit mock data is used to test whether the tension is driven by any single dataset. By replacing each real data with the mock generated one, we effectively remove its contribution to the shift of the neutrino mass posterior mean away from the $\sum m_\nu=0.058~\mathrm{eV}$.

\begin{figure}
    \centering
    \includegraphics[width=0.95\linewidth]{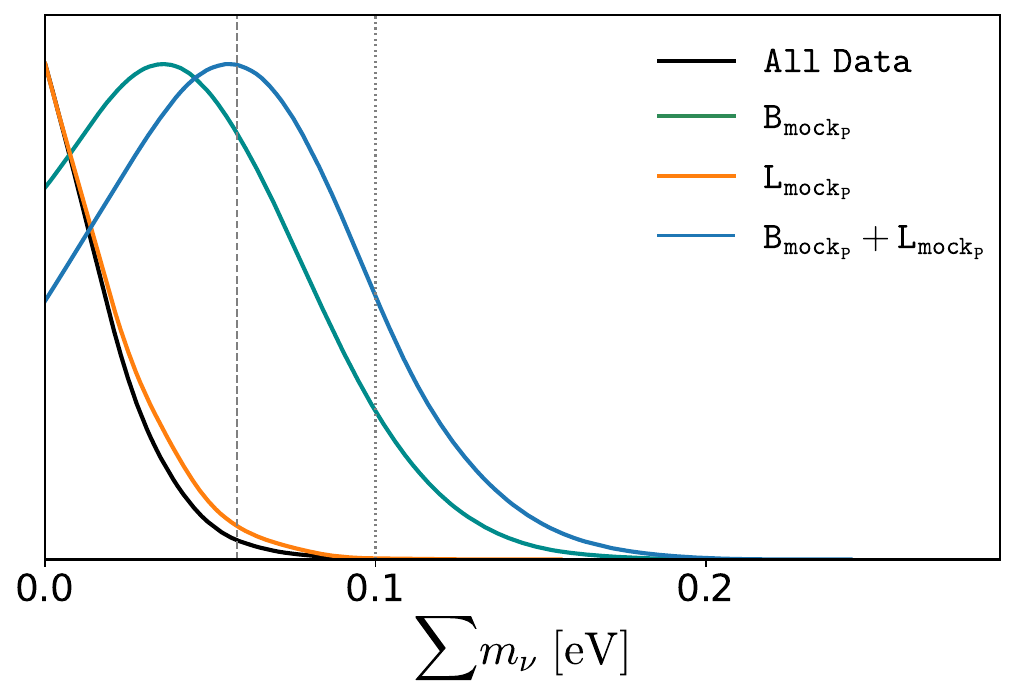}
    \includegraphics[width=0.95\linewidth]{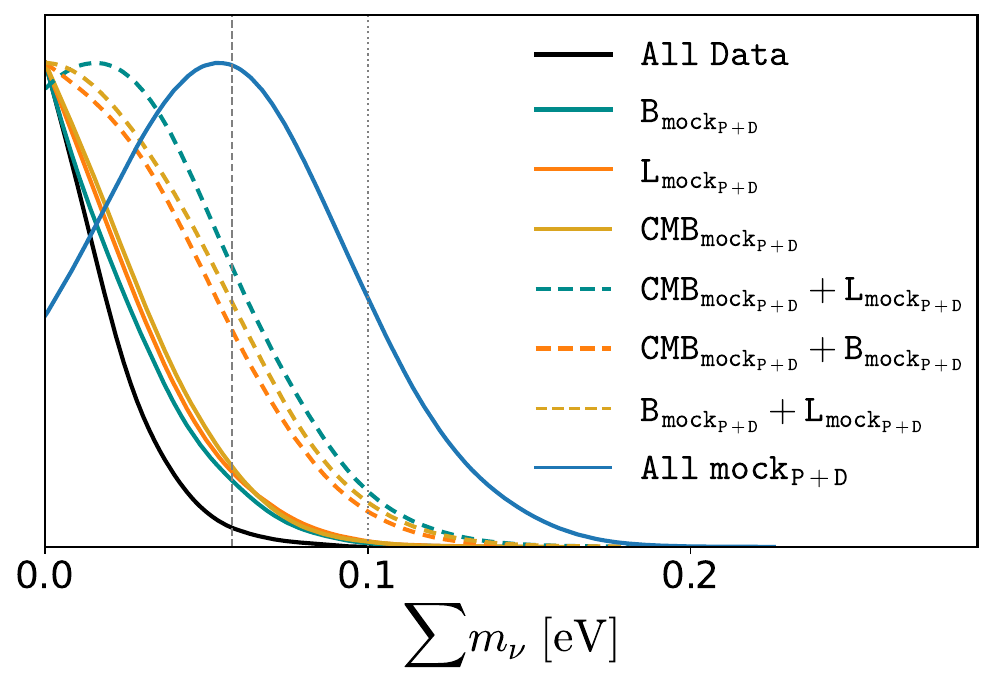}
    \caption{Constraints on $\sum m_\nu$ when the data vector of 
    various likelihoods are replaced
    with predictions from a fiducial $\Lambda$CDM cosmology, suggesting that no single dataset drives the tight upper bound. 
    We use shorthand \texttt{B} and \texttt{L} for mock BAO and CMB lensing likelihood respectively.
    The vertical dashed and dotted lines mark the minimum neutrino mass sum from the normal and inverted hierarchies, at 0.058 eV and 0.1 eV respectively. 
    In the top panel, BAO and/or CMB lensing data are replaced with mock data based on the best fit Planck \texttt{CamSpec} cosmology ($\text{mock}_\text{P}$).
    In the bottom panel, we instead replace the CMB, CMB lensing and/or BAO data with mock data based on the best fit Planck \texttt{CamSpec} + DESI cosmology ($\text{mock}_{\text{P}+\text{D}}$).
    In both cases, the $\sum m_\nu$ posterior is brought in better agreement with oscillation bounds.
    This suggests that the preference for low $\sum m_\nu$ is driven by a disagreement between the preferred $\Lambda$CDM parameters from Planck and DESI, rather than a disagreement between $\Lambda$CDM and each of these experiments separately. 
    } 
    \label{fig:fiducial}
\end{figure}

The top panel of Fig.~\ref{fig:fiducial} shows the difference in the constraints on $\sum m_\nu$ when replacing measured datasets with predictions from the \texttt{CMB}-only best-fit cosmology, while the baseline results from the actual measurements are included as a reference. We start with the constraints inferred from the combination of \texttt{CMB}, \texttt{Lensing} and \texttt{BAO}. The constraint on the neutrino mass sum is
\begin{equation}\label{eq:mnu2}
     \sum m_\nu < 0.0597 \ \mathrm{eV \ (95\% \ \mathrm{C.L.})}
\end{equation}
which is labelled as \texttt{All Data} in Fig.~\ref{fig:fiducial}. This excludes the minimum neutrino mass sum of the inverted hierarchy at the $>3\sigma$ level, and is close to the minimum value of the normal hierarchy from the neutrino oscillation experiments. This upper bound also tightens to $\sum m_\nu < 0.055\ \mathrm{eV \ (95\% \ \mathrm{C.L.})}$  when one combines the CMB primary data from \textit{Planck}, ACT DR6 and SPT-3G D1\citep{camphuis2025spt3gd1cmbtemperature}.

We see that substituting the \texttt{default} \texttt{BAO} likelihood with \texttt{mock} \texttt{BAO} likelihood set at the \texttt{CMB}-only bestfit largely relaxes the constraint on $\sum m_\nu$ (see the teal line in the top panel of Fig.~\ref{fig:fiducial}). 
The posterior distribution peaks at a positive value, slightly smaller than the fiducial value of 0.058 eV. The remaining difference is caused by the CMB lensing data. 
When we further replace the \texttt{default Lensing} likelihood with the \texttt{mock Lensing} likelihood set at the \texttt{CMB}-only bestfit (the blue curve in the top panel of Fig.~\ref{fig:fiducial}), the maximum posterior point of $\sum m_\nu$ returns to around the mock input value at 0.058 eV. 
This indicates that the leading cause of the tight $\mnu$ constraint is the disagreement between \texttt{default} \texttt{CMB} and \texttt{BAO}, while the \texttt{Lensing} likelihood plays only a sub-dominant role. This is further shown when only replacing the \texttt{default lensing} likelihood with \texttt{mock CMB}-only \texttt{Lensing} likelihood from the \texttt{All Data} combination, which results in a minor difference in the constraint on $\sum m_\nu$ (see the orange and black lines in the top panel of Fig.~\ref{fig:fiducial}).

This is in agreement with the results of \citep{Loverde:2024nfi, lynch2025whatsmattersigmamnu}, since the $\Omega_m$ constraint in the mock BAO data is in perfect agreement with that inferred from the primary CMB and CMB lensing data. However, given that there is a mild tension between BAO and CMB in $\Lambda$CDM, especially in the constraint on $\Omega_m$, the tight constraint on the upper bound of $\mnu$ is a result of the mild disagreement between CMB and BAO data.

To further illustrate the point that the tight upper bound of $\mnu$ is due to the mild disagreement between BAO and CMB data, we repeat the analysis using \texttt{mock} likelihood from the joint \texttt{CMB+BAO} best-fit cosmology, without putting a strong prior on CMB measurements being the truth. As shown in the bottom panel of Fig.~\ref{fig:fiducial}, the significant impact of BAO measurements goes away. Replacing any one of \texttt{BAO}, \texttt{CMB} or \texttt{Lensing} measurements with the \texttt{mock CMB+BAO} best-fit cosmology causes only small differences in the constraint on $\sum m_\nu$, indicating that none of the datasets is driving the neutrino mass tension in this case. When replacing two of \texttt{default} \texttt{BAO}, \texttt{Lensing} or \texttt{CMB} likelihoods with \texttt{mock CMB+BAO} likelihoods, we find that the center of the $\sum m_\nu$ posterior distribution shifts toward a positive value. The $\sum m_\nu$ posterior peaks at the value of 0.058 eV when we use the \texttt{mock CMB+BAO} likelihoods for all three cosmological probes.

To summarize, we show the shift of the $\mnu$ posterior towards negative values due to disagreement among \texttt{CMB}, \texttt{BAO} and \texttt{Lensing} data sets. The leading cause is the disagreement of \texttt{CMB} with \texttt{BAO}, which is still slightly below $3\sigma$ level. The source of the disagreement remains unknown, which could be statistical fluctuation, new physics or unresolved systematics biasing the results of either observations, but it is below the threshold forbidding combining the \texttt{CMB} and \texttt{BAO} likelihood to constrain $\Lambda$CDM parameters. If the true cosmology is the \texttt{CMB+BAO} best fit, no single data is the definite driver of the tight $\sum m_\nu$ upper limit. Thus, it is not informative to split the $\sum m_\nu$ based on data. 
Although \texttt{BAO} contains only geometrical information of the late Universe, \texttt{CMB} contains both geometry and growth information. Splitting $\sum m_\nu$ by data is not able to answer the question of which physical channel drives the $\sum m_\nu$ anomaly, geometry or growth. 
Thus, in Sect.~\ref{sec:split_imp}, we introduce and implement a phenomenological split of $\mnu$ into geometrical and growth channels, namely $\mgeo$ and $\mgrowth$. 
The shifts of $\mnu$ posterior mean observed in this section also motivates us to look into the \texttt{mock} data and fisher forecast before using the real data in Sect.~\ref{sec:split_results}.

\section{Literature review and how our approach differs}\label{sec:review}

Before detailing the implementation of splitting $\mnu$ into geometry part and growth part in Sect.~\ref{sec:split_imp}, we briefly summarize and review previous work trying to understand the source of tight $\sum m_\nu$ constraint.

\subsection{{Preference for a lower energy density at low redshift}}

Several studies have focused on understanding the role of background expansion history and the overall matter density $\Omega_m$ in driving the negative neutrino mass preference. In \citep{Loverde:2024nfi, elbers2025constraintsneutrinophysicsdesi}, the authors point out the sensitivity of  the neutrino mass constraint to the geometrical information of the cosmological datasets. 
The geometrical information is at least as important as the structure growth effects (identified with CMB lensing) in constraining $\mnu$. The geometrical information at low redshifts, from BAO and SNe, is sensitive to the overall matter density, while the high-redshift sound horizon inferred from CMB is also sensitive to the CDM and baryon density. Thus, the neutrino mass sum is inferred by the difference between the total mass in the Universe and the sum of CDM and baryons. The tight constraint on neutrino mass when combining CMB with DESI DR1 
is due to a low total mass fraction at present preferred by BAO contrasting to the CDM and baryon density inferred from CMB.

\subsection{Preferences for ``negative'' values of neutrino masses, without splitting geometry and growth}

In \citep{Elbers:2024sha, elbers2025constraintsneutrinophysicsdesi}, the authors consider the impacts of massive neutrinos on the background geometrical information and the growth of the structure simultaneously by extrapolating all cosmological observables to negative neutrino mass by 
\begin{equation}
    X\left(\sum {\tilde{m}_\nu} \right)= X_0 
    + \mathrm{Sgn}(\sum \tilde{m}_\nu)
    \left[X\left(\abs{\sum{\tilde{m}_\nu}} \right)
    - X_0\right],
\end{equation}
where $X_0 = X\left(\sum {{m}_\nu} = 0 \right)$, Sgn is the same sign function defined as
\begin{equation}\label{eqn:sign}
{\rm Sgn}(\sum\tilde{m}_\nu) = \begin{cases} +1, & \sum\tilde{m}_\nu \geq 0 \\ -1, & \sum\tilde{m}_\nu < 0 \end{cases},
\end{equation}and $X$ represents all cosmological observables, including primary CMB, CMB lensing power spectra, and BAO. They found that the minimum neutrino mass sum of NO hierarchy is in 3$\sigma$ tension \cite{elbers2025constraintsneutrinophysicsdesi}. They attribute the source of the tension to a higher $H_0r_d$ preferred by DESI DR2 BAO than the one inferred from CMB.

\subsection{Anomalous expansion history, at fixed growth}\label{sec:background_negmnu}

In \citep{lynch2025whatsmattersigmamnu}, the authors quantify the matter-deficiency that causes the anomalous $\mnu$ upper bound by building a background-only likelihood using a ``signed neutrino mass'' model. In this model, the expansion rate, $H(z)$, is anchored around $\sum m_\nu=0$ symmetrically with
\begin{align}
    H^2(z, \sum \bar{m}_\nu) = & H^2(z, \sum m_\nu=0)+\mathrm{Sgn}(\sum \bar{m}_\nu)\times \nonumber \\
    &\left[H^2(z, |\sum \bar{m}_\nu|)-H^2(z, \sum m_\nu=0)\right], \nonumber
\end{align}
where Sgn is defined in Eqn.~(\ref{eqn:sign}).
The background-only likelihood is then built by compressing the full primary CMB analyses to Gaussian approximations of the marginal posteriors $P(\omega_b, \omega_c, \theta_s^*)$ assuming $\Lambda$CDM model with $\sum m_\nu=0.06$eV, which encapsulates the geometry information from the primary CMB. Combining with DESI BAO data, which constitute purely geometrical information at low redshifts, they found $\sum \bar{m}_\nu = (-0.123 \pm 0.081)$\,eV, a 2.3$\sigma$ preference for a matter density deficit relative to the case of the minimum NO expectation of 0.06 eV, when using the background-only CMB likelihood derived from \textit{Planck} PR3 likelihoods. The result shows that the background expansion history alone, through the interaction of BAO and CMB distance constraints, drives a preference for negative neutrino masses independent of lensing considerations. 
However, this choice is not completely free from growth information, since $P(\omega_b, \omega_c, \theta_s^*)$ is marginal posteriors derived from the full CMB likelihood.
The $\sum \bar{m}_\nu$ is shifted to ($-0.017 \pm 0.089$)\,eV, which is $\sim \ 0.9\sigma$ lower than 0.06\,eV, when switching to a background-only likelihood derived from $\Lambda$CDM+$A_\mathrm{2pt}$ model on the same \textit{Planck} PR3 primary CMB likelihood, in which freeing up $A_\mathrm{2pt}$ allows the growth information to deviate from the fiducial fixing $\sum m_\nu=0.06\,$eV case.

\subsection{Anomalous growth, at fixed expansion}

Other studies have examined the role of CMB lensing and structure growth in the tension, with more recent work attempting to disentangle these effects from the background expansion.  The authors of \citep{craig2024nusgoodnews,PhysRevD.111.083507} analyzed the negative neutrino mass preference caused by the excess lensing effect by using a parametrization where deviations from $\Lambda$CDM manifest as changes to the CMB lensing effects. The CMB lensing potential is extrapolated based on $\sum m_\nu$ around the fiducial predictions, and this lensing potential is used to produce the lensed primary CMB spectra computed assuming $\sum m_\nu=0$.
However, in the same model with $\sum m_\nu=0$ for the unlensed primary CMB spectra, the BAO observables are computed. They found that CMB lensing drives the preference for negative neutrino mass, and that this preference remains robust to variations in optical depth measurements, BAO datasets, and assumptions about late-time dark energy evolution. However, their parametrization modifies the lensing amplitude (affecting $C_L^{\phi\phi}$ and its impact on the two-point CMB power spectra through peak smearing and other effects) while holding the expansion history fixed, thereby not self-consistently accounting for how modifications to the neutrino sector affect both clustering and distances simultaneously. 

\subsection{Anomalous growth, marginalized expansion}

In \citep{cozzumbo2025shortblanketcosmologycmb}, the authors use the same parametrization as \citep{elbers2025constraintsneutrinophysicsdesi} and include additional parameters to scale the amplitude of the CMB lensing potential directly and the amount of lensing effects in the lensed primary CMB power spectra. They test different high-$\ell$ and low-$\ell$ CMB likelihood choices in combination with the DESI DR2 BAO likelihood and the \texttt{P-ACT} CMB lensing likelihood. They find that it is possible to resolve the neutrino mass tension in the extended model, with the price of a CMB lensing amplitude slightly higher than 1. 
They thus conclude that the main culprit for the neutrino mass tension is CMB lensing being too high compared to the primary CMB prediction.

\subsection{Splitting the neutrino mass parameter based on observable rather than effect}

The analysis of \citep{graham2025darkforcesgathering} introduces two independent parameters, $\sum m_\nu^{\rm clustering}$ and $\sum m_\nu^{\rm BAO}$, which separately capture the effects on CMB lensing potential and the expansion rate measured from BAO observables. Their approach reveals that while both parameters individually prefer negative values when the other is held at zero, the degeneracy between them shows that new physics to either channel alone could resolve the current tension. This two-dimensional framework demonstrates that modifications to expansion history (captured by $\sum m_\nu^{\rm BAO}$) can shift the expected lensing signal in a way that allows for positive $\sum m_\nu^{\rm clustering}$, and vice versa. Although BAO is a purely geometry observable, it only capture the expansion rate at the late Universe. Parametrizing only as $\sum m_\nu^{\rm BAO}$ ignores the geometrical information contained in the CMB observables. 

\subsection{Splitting the neutrino mass parameter based on background and neutrino-only perturbations}

In \cite{Bertolez-Martinez:2024wez}, the authors studied the constraints on neutrino mass sourced from its impact on the expansion rate via neutrino energy density and from the neutrino mass perturbation evolution via the ``kinematic'' free-streaming effects, by introducing two effective masses, $\sum m_\nu^{\rm Backg.}$ and $\sum m_\nu^{\rm Pert.}$, respectively. 
Note, however, that here $\sum m_\nu^{\rm Pert.}$ does not capture the effect of the neutrino energy density on the matter and photon perturbations, only the effect of the neutrino perturbations themselves, which is expected to be small beyond a phase shift of the CMB BAO peaks.
Constrained with {\it Planck} 2018 PR3 CMB dataset, they indeed find that the neutrino mass is mostly constrained through its change to the expansion rate in the background cosmology. 
As expected, the constraints inferred from the neutrino perturbation is weak, with 95\% C.L. upper bound on $\sum m_\nu^{\rm Pert.}$ approximately 4 times of that in $\sum m_\nu^{\rm Backg.}$. 
However, given that the structure formation is mainly sourced by the cold dark matter and that the fraction of neutrinos in the total energy is very small, separating neutrino perturbation alone from the background expansion still mixes the effect of the massive neutrino energy density on the geometry and growth of matter fluctuations.

\subsection{Possible resolution with a different CMB optical depth $\tau$, curvature, dynamical dark energy}\label{sec:param_deg}

In \citep{Sailer:2025lxj,Jhaveri:2025neg,Sullivan:2026tas,Allali:2025yvp}, the authors explored the possibility  of alleviating the ``negative neutrino mass'' issue by relaxing the constraints on the reionization optical depth. The reionization optical depth is the least constrained parameter in the $\Lambda$CDM model, and relies almost completely on the \textit{Planck} low-$\ell$ EE likelihoods. By relaxing the $\tau$ constraint, it is possible to explore the degeneracy between $\tau - A_s - \omega_c - \sum m_\nu$ directions. The preferred $\tau$ to recover the minimum neutrino mass sum is $\sim 0.09$, while $\tau \sim 0.054$ from \textit{Planck} low-$\ell$ likelihoods.

In terms of subtle model dependence, apart from suggestions that hints of dynamical dark energy relax the neutrino bounds (see e.g.~\citep{Shao:2024mag, desicollaboration2025desidr2resultsii, elbers2025constraintsneutrinophysicsdesi}), authors of \citep{chen2025itsttokcurvature} explored spatial curvature as a physical resolution to this tension, finding that DESI DR2 BAO combined with CMB data prefers a $2\sigma$ detection of negative spatial curvature with $\Omega_k = 0.0023 \pm 0.0011$, corresponding to a comoving curvature radius of $R_k \approx 21H_0^{-1}$. Allowing for nonzero spatial curvature significantly relaxes cosmological constraints on the sum of neutrino masses, with the upper limit expanding from $\sum m_\nu < 0.062$ eV to $\sum m_\nu < 0.10$ eV at $95\%$ confidence when curvature is treated as a free parameter. This relaxation occurs because both curvature and neutrino mass break the same geometric degeneracy in the angular acoustic scale $\theta_\star$ measured by the CMB, following approximately $1.4\Omega_k \approx 0.4f_\nu$. This degeneracy is steep enough that both the normal and inverted neutrino mass hierarchies become compatible with current data when spatial curvature is allowed to vary, in contrast to the preference for sub-minimal masses in flat $\Lambda$CDM models.

Although extensions of dynamical dark energy and curvature can alleviate the neutrino mass tension, additional parameters degenerate with $\sum m_\nu$ are introduced to absorb the tension between CMB and BAO. 
Since the same data combination also shows preference for dynamical dark energy, especially for the BAO bins at low redshift, authors of \cite{Qu:2026fkt} seek to find a channel whose constraints on $\sum m_\nu$ are independent of the low-redshift regime. 
The ``late-Universe-free'' combination consists of the primary CMB with $A_\mathrm{lens}$ marginalization (namely the ``unlensed'' CMB) and the reconstructed CMB lensing potential power spectrum ($C_L^{\kappa\kappa}$), identified by inspecting the CMB and BAO data probe by probe. Encountered with current CMB and BAO data, this ``late-Universe-free'' channel gives $\sum m_\nu < 0.41~\mathrm{eV} ~\mathrm{(95\% C.L.)}$. When the dark energy equation of state $w(a)$ is also marginalized, the ``late-Universe-free'' channel produces similar $\sum m_\nu$ constraints, indicating that such a channel does not include the disputed low-redshift information.

\section{Expansion VS growth: impact on observables, and constraints on $\mgeo$, $\mgrowth$}\label{sec:split_imp}

For the neutrino mass range relevant to this work, the transition happens during the matter dominated era. Massive neutrino impacts both the geometry and structure growth of the Universe after its transition from relativistic to non-relativistic by adding unclustered energy density to the total matter.
Such a transition boosts the cosmic expansion rate. The low-redshift angular observables, like BAO, are affected by such a change in the expansion rate due to the change in the angular diameter distance. Due to the free streaming of relativistic neutrinos, this transition reduces the small-scale clustering after the neutrinos turn non-relativistic. Also, a faster expansion rate slows down the structure growth of all components of the Universe. 

It is possible to relieve the tight $\sum m_\nu$ constraints by exploiting the parameter degeneracy directions as discussed in Sect.~\ref{sec:param_deg}. The solutions with reionization optical depth $\tau$ open up the degeneracy with $\sum m_\nu$ in the structure growth channel, while the addition of curvature and dynamical dark energy opens up the geometry channel. The question of whether the geometry or growth channel favors a smaller neutrino mass sum remains unanswered. As discussed in Sect.~\ref{sec:review}, other works did not separate the information from geometry and growth coherently. Thus, in this section, we propose and implement a proper phenomenological split of the geometry and growth effect from massive neutrinos.

\subsection{Phenomenologically splitting the effect of geometry and growth}

In this section, we introduce the phenomenological split of massive neutrino effects into geometry and growth parts, parametrized with $\mgeo$ and $\mgrowth$ respectively. We will briefly review the computation of CMB power spectra and BAO observables before defining the split. 

\begin{figure*}
    \centering
    \includegraphics[width=0.98\linewidth]{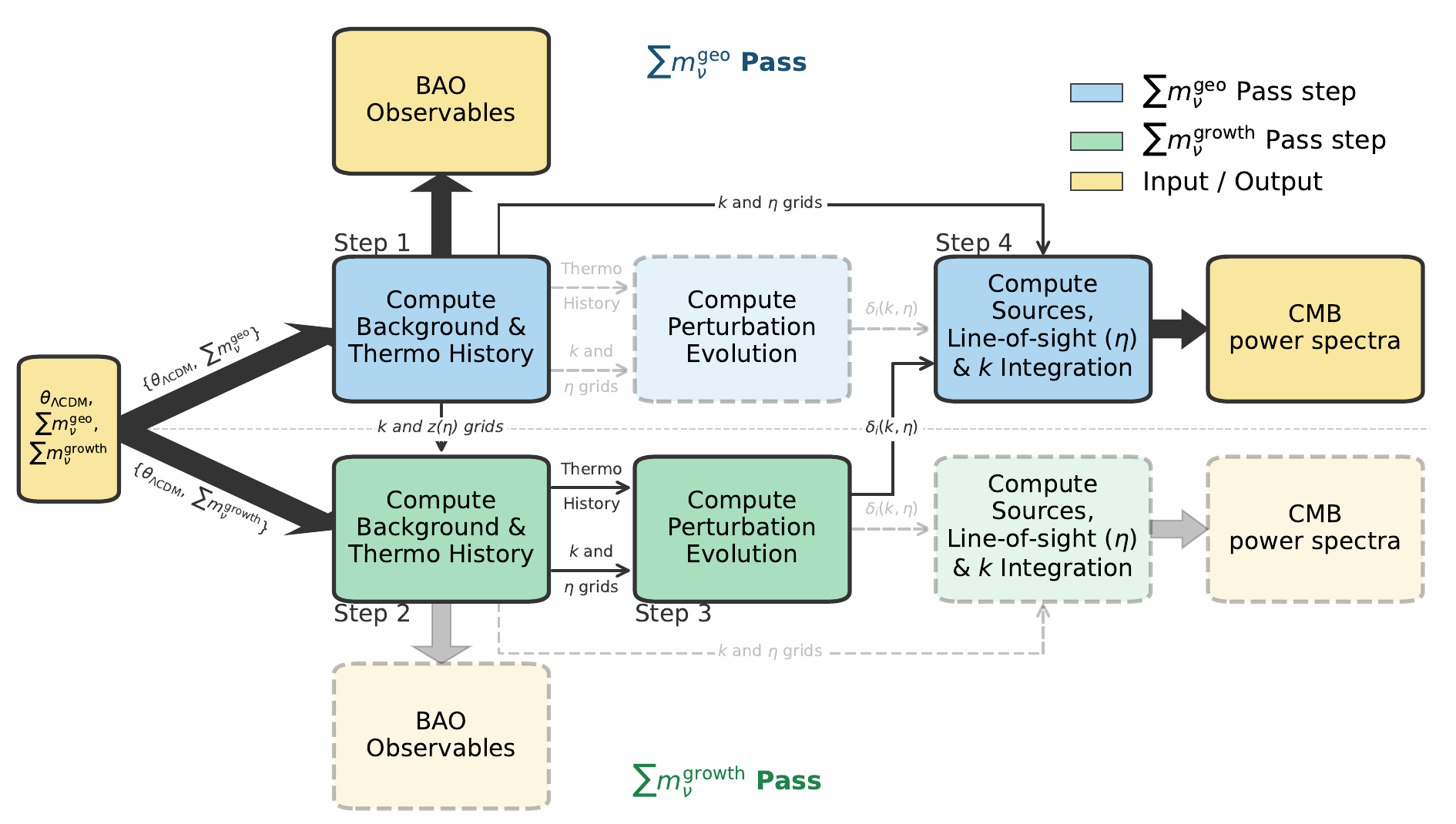}
    \caption{Flow chart of the implementation of $\mgeo$ and $\mgrowth$ in \texttt{CAMB}. The dashed gray line in the middle separates two independent \texttt{CAMB} runs; our modification for the $\theta_{\Lambda\rm{CDM}} + \mgeo + \mgrowth$ model is implemented by stitching them together. We set $\theta_\mathrm{\Lambda CDM} = \{H_0, \omega_b, \omega_c, A_s, n_s, \tau\}$. The yellow boxes represent inputs and outputs; the blue and green boxes represent the computations in the two independent \texttt{CAMB} calls before our modifications. The dark boxes and arrows in solid black lines are used in the $\theta_{\Lambda\rm{CDM}} + \mgeo + \mgrowth$ case, while the light boxes and arrows with dashed lines are the skipped steps. Steps 1--4 correspond to the steps described in Sect.~\ref{sec:split_imp}. The grayed-out BAO output from the $\mgrowth$ pass is shown for symmetry with the standard \texttt{CAMB} workflow but is not used in our analysis, since BAO observables only depend on the geometry sector.
    }
    \label{fig:flowchart}
\end{figure*}

\textbf{CMB power spectrum} --- The CMB power spectrum can be expressed as
\begin{align} \label{eq:cls}
    C_\ell^{XY} \propto \int k^2 dk\, \mathcal{P}_\mathcal{R}(k)
    &\left[\int_0^{\eta_0}d\eta\, S^X(\eta, k) j_\ell(x)\right]\nonumber\\
    &\left[\int_0^{\eta_0}d\eta\, S^Y(\eta, k) j_\ell(x)\right],
\end{align}
where $\{X, Y\}\in\{T,E,\phi\}$, $x = k(\eta_0 - \eta)$, $\eta$ is the conformal time (or the comoving distance, because $\eta$ is proportional to the comoving distance $\chi$), and $S$ is the source term for the $T$, $E$, or $\phi$ signal \citep{Zaldarriaga:1996xe, Lewis:2006fu}. In the temperature source, there is no explicit functional form of $\eta$, where $\eta$ is only a time variable to the terms, like the monopole, dipole and ISW effects, showing up in the temperature source; however, the comoving distance appears directly in $S^E$ and $S^\phi$, instead of being a time variable for the quantities in the source terms. The CMB power spectra contain both geometry and growth information: the source term contains the perturbation quantities, imprinting the information from growth; while the geometry information enters through the functional form of $\eta$ in the polarization and lensing source, as well as the spherical Bessel function $j_\ell(x)$.

\textbf{BAO} --- The BAO measures the imprint of the baryon acoustic oscillation before matter-photon decoupling. Specifically, the BAO measures the ratio $r_d/D_H(z)$ in the line-of-sight direction, the transverse ratio $r_d/D_M(z)$, and the isotropic ratio $r_d/D_V(z)$, where $D_M(z)$ is the angular diameter distance to redshift $z$, $D_H(z) = c/H(z)$, and $D_V(z) = z D_M^2(z) D_H(z)$. It contains purely geometry information as it measures the angular baryon acoustic scales before the last scattering expanded on the sky. 

In our phenomenological model separating the geometry and growth information of massive neutrinos, we propose to extend the 6 free parameters of the $\Lambda$CDM model with two additional parameters, $\mgeo$ and $\mgrowth$, by requiring:
\begin{itemize}
    \item $\mgeo$ affects the expansion rate used to calculate the geometrical quantities. This includes the angular diameter distance and $D_H(z)$ for BAO, as well as the geometric factor in the source term of the CMB power spectra.
    \item $\mgrowth$ enters the expansion rate for the computation of the growth of fluctuations, including the perturbation evolution of all species in the Universe and the metric perturbations $\phi$ and $\psi$.
\end{itemize}
In practice, we ``stitch'' two cosmologies with expansion rates sourced from $\mgeo$ and $\mgrowth$ separately. This is implemented in a modified version of \texttt{CAMB}\footnote{The code is publicly available at \url{https://github.com/shuimufge/CAMB_geo_growth.git}}, and the steps are shown schematically in Fig.~\ref{fig:flowchart}:
\begin{enumerate}\label{step}
    \item Compute the background cosmology and the thermal history with \{$\theta_{\Lambda\rm{CDM}} + \mgeo$\}; save the grids of wavenumber $k$ and conformal time $\eta^{\rm geo}$. The BAO observables are derived from the background and thermal history computed in this step. \label{step:back_geo}
    \item Compute the background cosmology and the thermal history with \{$\theta_{\Lambda\rm{CDM}} + \mgrowth$\}; set the comoving $k$-grid for the perturbation evolution to be the same as the $k$-grid from Step~\ref{step:back_geo}, and create the conformal time grid $\eta^{\rm growth}$ by matching with the $\eta^{\rm geo}$ grid through redshift by requiring $z(\eta^{\rm geo}_i) = z(\eta^{\rm growth}_i)$, where $i$ indexes the conformal time points. 
    \label{step:pert_geo}
    \item Compute the perturbation evolution using the setup from Step~\ref{step:pert_geo}. \label{step:pert_growth}
    \item Compute the sources, the line-of-sight integration, and the $k$-integration for the CMB spectra by combining the geometry from Step~\ref{step:back_geo} with the perturbation quantities from Step~\ref{step:pert_growth}. The non-linear correction to the CMB lensing kernel using HMCode is also performed in this step by computing the non-linear ratio using the cosmology of $\mgrowth$.
\end{enumerate}

Now we discuss the choices we made in the above steps. We have chosen $\theta_\mathrm{\Lambda CDM}$ to be the standard 6 parameters with $\theta_\mathrm{\Lambda CDM} = \{H_0, \omega_b, \omega_c, A_s, n_s, \tau\}$. 
To implement the split of $\mgeo$ and $\mgrowth$, we effectively substitute the perturbations of overdensities ($\delta_i$ and $\theta_i$) and metric perturbations ($\psi$ and $\phi$ for Newtonian Gauge for example) in computing CMB power spectra in the $\mgeo$ pass with those computed in the $\mgrowth$ pass. 

The choice is not unique. For example, to get the CMB power spectra, one could compute the angular diameter distance to the last scattering surface ($D_A(z_*)$) from $\{\theta_\mathrm{\Lambda\mathrm{CDM}}, \mgeo\}$, then horizontally shift the CMB power spectra computed from $\{\theta_\mathrm{\Lambda\mathrm{CDM}}, \mgrowth\}$ based on the ratio of $D_A(z_*)$'s computed from the two cosmology. However, this is only an approximated results given the mismatch of between peak locations of the simple expectations and numerical results \citep{Pan:2016zla}. Our choice avoids this problem by conducting numerical computation based on the geometry and time evolution of the background of $\{\theta_\mathrm{\Lambda\mathrm{CDM}}, \mgeo\}$ directly. Thus, it is natural to implement the model with $\mgeo - \mgrowth$ extension as our choice. 

Now we discuss the choices and the subtleties involved in setting the $(k, \eta)$ grids for the two cases.
For the $k$ grid, we use the exact comoving $k$ grids for computing the perturbation evolution pass as one would use in the complete $\mgeo$ pass. The comoving $k$ grids are composed of a mixture of log spaced points for mode outside the horizon and linearly spaced points inside the horizon. 
For the conformal time grid, we choose to match the $\eta^{geo}$ grid with the $\eta^{growth}$ grid by fixing the corresponding redshift grids of the two cases. The ages of the $\{\theta_\mathrm{\Lambda\mathrm{CDM}}, \mgeo\}$ and $\{\theta_\mathrm{\Lambda\mathrm{CDM}}, \mgrowth\}$ cosmologies are not necessarily the same, which will lead to numerical problems in the line-of-sight integration of Eqn.~\ref{eq:cls}. Thus, we connect the $\eta$ grids of the $\mgeo$ and $\mgrowth$ cosmologies through redshift. We first convert the $\eta^{geo}$ grid to $z$ grid under the \{$\theta_{\Lambda\rm{CDM}} + \mgeo$\} cosmology, then we compute the $\eta^{growth}$ grid of the same $z$ grid using the \{$\theta_{\Lambda\rm{CDM}} + \mgrowth$\} cosmology. 

The choice to match the $(k,z)$ grid here is physically motivated. The $k$'s are comoving wavenumbers, which are also related to the geometric information via $k\sim \ell/\chi_0^{geo}$ in CMB, so it is natural to match the $k$ grid in $\{\theta_\mathrm{\Lambda\mathrm{CDM}}, \mgrowth\}$ with $\{\theta_\mathrm{\Lambda\mathrm{CDM}}, \mgeo\}$. To see the choice for matching the $\eta$ grids between two cosmologies, we perform a change of variable of the line-of-sight integral in the square brackets in Eqn.~\ref{eq:cls}:
$$
    \int_0^{\eta_0}d\eta\, S(\eta, k) j_\ell(x)=\int_0^{\infty}\frac{dz}{H(z)}\, S[z, k] j_\ell(x).
$$
Since the two cosmologies share the same \{$\theta_\mathrm{\Lambda\mathrm{CDM}}$\} and neutrinos remain relativistic at the early time, the early-time physics are the same in both cases.
The shared physical scales imprinted during recombination, as well as the normality of the visibility function at early universe, are automatically preserved. At late times when the massive neutrinos become non-relativistic, the expansion rates of the two cosmologies differ, leading to different conformal age $\eta_0$. If matching the conformal time directly instead of matching the redshift, the integration interval will have mismatch. As a result, when the age of the Universe in $\mgrowth$ pass is younger than that in $\mgeo$ pass, the integrand will needs to be extrapolated. However, the bounds of the redshift integration are always the same. 

The matching by redshift is different from the matching by conformal time used in the ``separate Universe'' approach of N-body simulations studying the impact of long-wavelength perturbations to small-scale observables within a finite survey or simulation box, where the long-wavelength overdensity can be absorbed into an alternative cosmology instead of increasing the simulation box or survey size (see e.g. \citep{Wagner:2014aka, Li:2014sga, Hu:2016ssz}). The requirement originates from the equal background matter density at any conformal time between the two cosmologies. However, in our split, we effectively split the expansion rate into geometry and growth channel via neutrino mass sum, which is effectively different from the separate Universe.

We note that any matching convention defining a valid phenomenological model should reduce to $\Lambda\mathrm{CDM}+\Sigma m_\nu$ when $\mgeo = \mgrowth$. We have compared the CMB power spectra computed by setting $\mgeo=\mgrowth=\tilde{m}_\nu$ with the one from unmodified \texttt{CAMB} by setting $\sum m_\nu = \tilde{m}_\nu$. The two agree within numerical precisions. 

\subsection{Impact of $\mgeo$ on BAO}

\begin{figure}[b]
    \centering
    \includegraphics[width=0.95\linewidth]{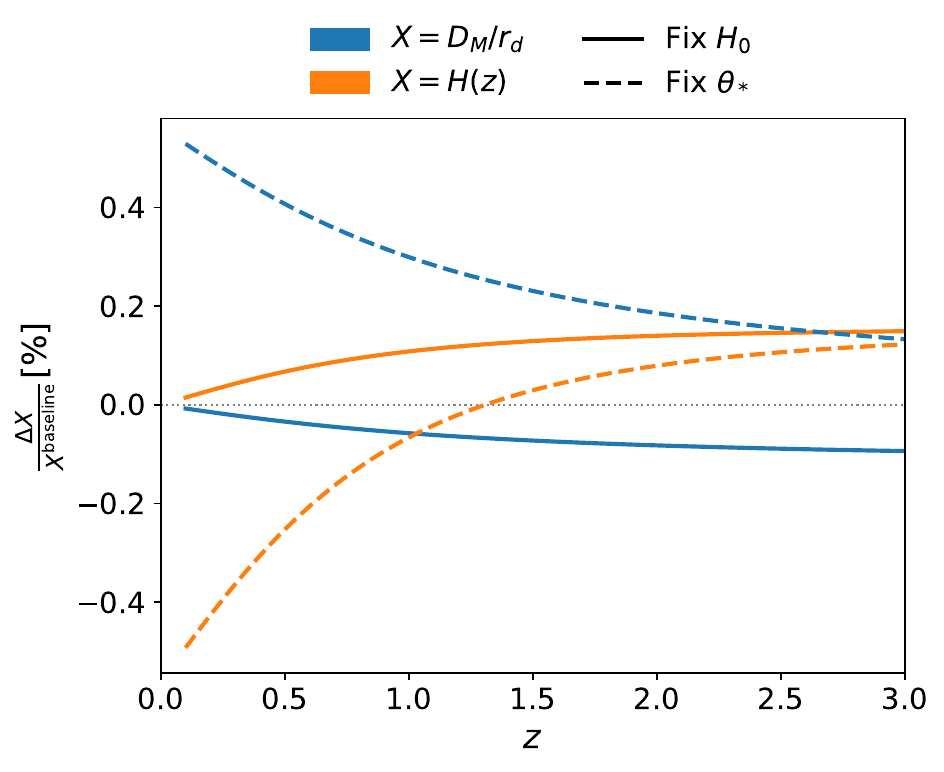}
    \caption{Fractional changes of $D_M(z)/r_d$ (in blue) and $H(z)$ (in orange) of the case with $\mgeo=0.1$eV compared to the baseline with $\mgeo=0.058$eV. The solid lines and the dashed lines show the choices of keeping $H_0$ and $\theta_*$ fixed respectively.}
    \label{fig:bao_mnu_geo}
\end{figure}

The transition from relativistic to non-relativistic of massive neutrinos alters the expansion rate at late time, thus it impacts the BAO observables. In Fig.~\ref{fig:bao_mnu_geo}, we show the fractional changes of $D_M/r_d$ and $H(z)$ as a function of redshift comparing $\mgeo=0.1$eV case to the baseline $\mgeo=0.058$eV case\footnote{Note that the BAO is a geometrical observable only, so it does not depend on $\mgrowth$.}. 
When holding $H_0$ fixed (solid lines in Fig.~\ref{fig:bao_mnu_geo}), increasing $\mgeo$ boosts the expansion rate during matter dominated era and delays the matter-DE equality. As a result, we see an increase in $H(z)$ compared to the baseline case, with a decreasing amount of excess as redshift approaching 0. The $D_M/r_d$ and $D_H/r_d$ observables of BAO decrease in this case, compared to the fiducial model, since $H(z)$ shows up in the denominator when computing the two observables. On the other hand, if we choose to fix $\theta_*$ instead, the fractional change of $H(z)$ crosses zero as a function of redshift. At matter dominated era, the increase of $\sum m_\nu$ boosts the total matter density, thus increasing the expansion rate. However, the angular diameter distance to the last scattering surface is fixed due to the choice of fixing $\theta_*$ and other cosmological parameters. The expansion rate after DE-domination has to slow down relative to the baseline case to keep $D_A(z_*)$ unchanged, resulting to decreased expansion rate at low-z when increasing $\sum m_\nu$ with $\theta_*$ fixed. As a result of the change to $H(z)$, $D_M/r_d$ and $D_H/r_d$ decrease with increasing redshift.

\subsection{Impact of $\mgeo$ on CMB and CMB lensing}
\label{sec:mnu_geo_cl}

\begin{figure*}[!]
    \centering
    \includegraphics[width=0.9\linewidth]{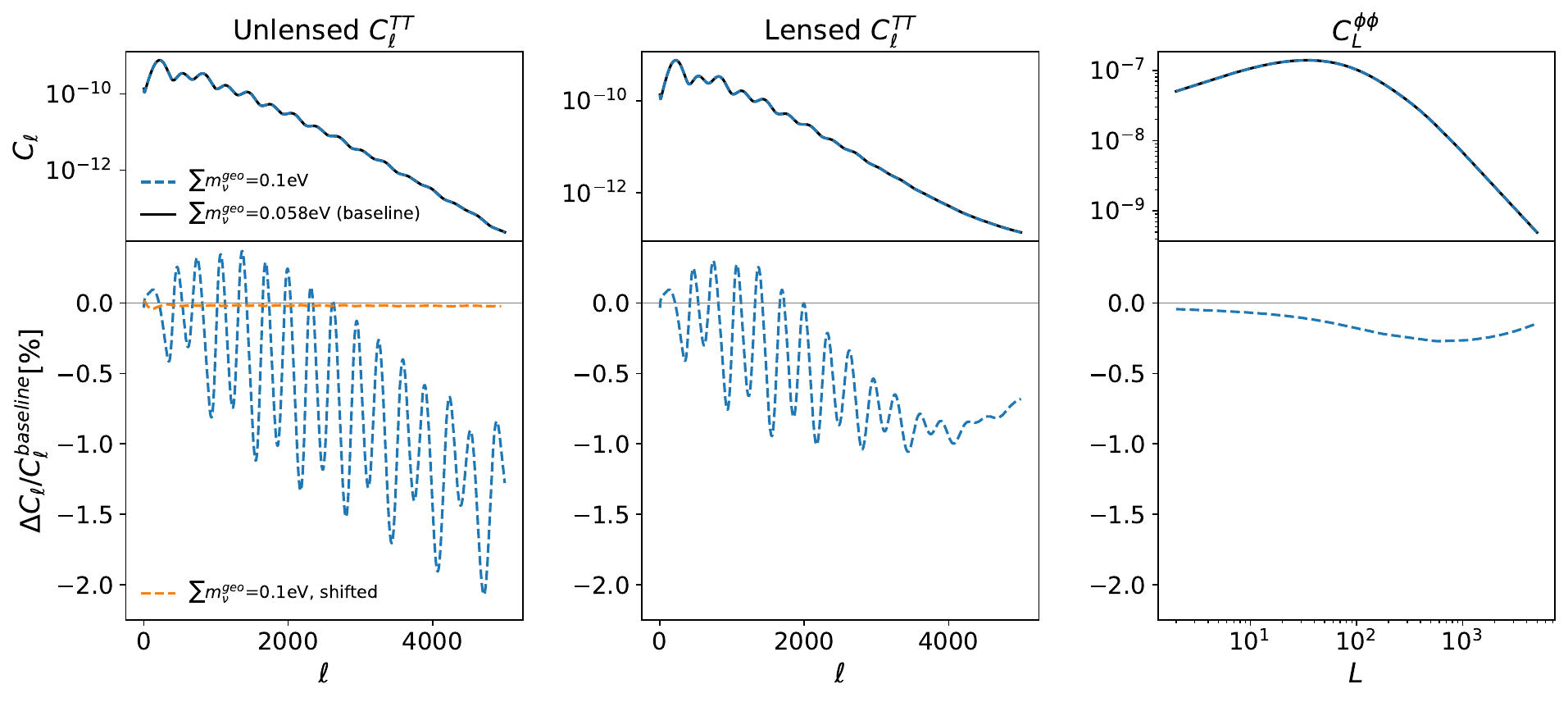}
    \caption{CMB temperature and lensing power spectra when changing $\mgeo$ with $\mgrowth$ held at 0.058~eV. The unlensed CMB TT spectrum is shown in the left panels and the lensed one is in the middle. The top panels show the power spectra and the bottom show the fractional difference to the base line result. The orange dashed line is the fractional difference of $\mgeo=0.1\mathrm{eV}$ after correcting for the phase shift due to the $\ell \sim k\cdot D_A$. On the right shows the CMB lensing potential power spectra. The baseline model is $\mgeo=0.058$eV.
    }
    \label{fig:geo_effect}
\end{figure*}

In Fig.~\ref{fig:geo_effect}, we show the changes to the CMB temperature and CMB lensing potential power spectra when changing $\mgeo$ with $\mgrowth=0.058$eV fixed. The geometry information is contained in the peak and trough locations of CMB primary power spectra. Changes in the background geometry will lead to phase shifts in the unlensed CMB power spectrum. The peak locations can be approximated by $\ell \sim kD_A(\mgeo)$ (see e.g.~\citep{Pan:2016zla}), where the angular diameter distance $D_A$ depends on $\mgeo$. As massive neutrinos become non-relativistic at late time, changing $\mgeo$ will alter the function of late time expansion rate, thus altering the angular diameter distance to the last scattering surface. Such a change is manifested as a phase shift in the power spectrum. Once we shift the unlensed power spectra based on the changes to $D_A$, the relative difference of changing $\mgeo$ is close to zero (see blue and orange lines in the bottom left panel of Fig.~\ref{fig:geo_effect}). We have verified a similar response of unlensed EE power spectrum to a change of $\mgeo$.

Changing $\mgeo$ has a smaller impact on CMB lensing potential. The fractional changes of increasing $\mgeo$ are shown in the bottom right panel of Fig.~\ref{fig:geo_effect}. Such a change is caused by a combination of the phase shift due to the change in $D_A$ as well as the response of the CMB lensing kernel. Although the $j_\ell(kx)$ factor also shows up in $C_L^{\phi\phi}$ (see Eqn.~\ref{eq:cls}), the lensing kernel has a different geometric window. As a result, the changes to $C_L^{\phi\phi}$ by $\mgeo$ are not exactly the same as that in $C_\ell^{TT}$.

Lastly, the changes to the lensed CMB temperature power spectrum are shown in the middle panel of Fig.~\ref{fig:geo_effect}. At large angular scales, where the CMB lensing effect is small, the fractional changes in the lensed $C_\ell^{TT}$ is similar to the changes in unlensed $C_\ell^{TT}$ caused by changing $\mgeo$. On small angular scales, where the CMB lensing effect transports power from low-$\ell$ to high-$\ell$, the pure phase shift effect is washed out by the lensing effect. 

\subsection{Impact of $\mgrowth$ on CMB and CMB lensing}
\label{sec:mnu_growth_cl}

\begin{figure*}
    \centering
    \includegraphics[width=0.9\linewidth]{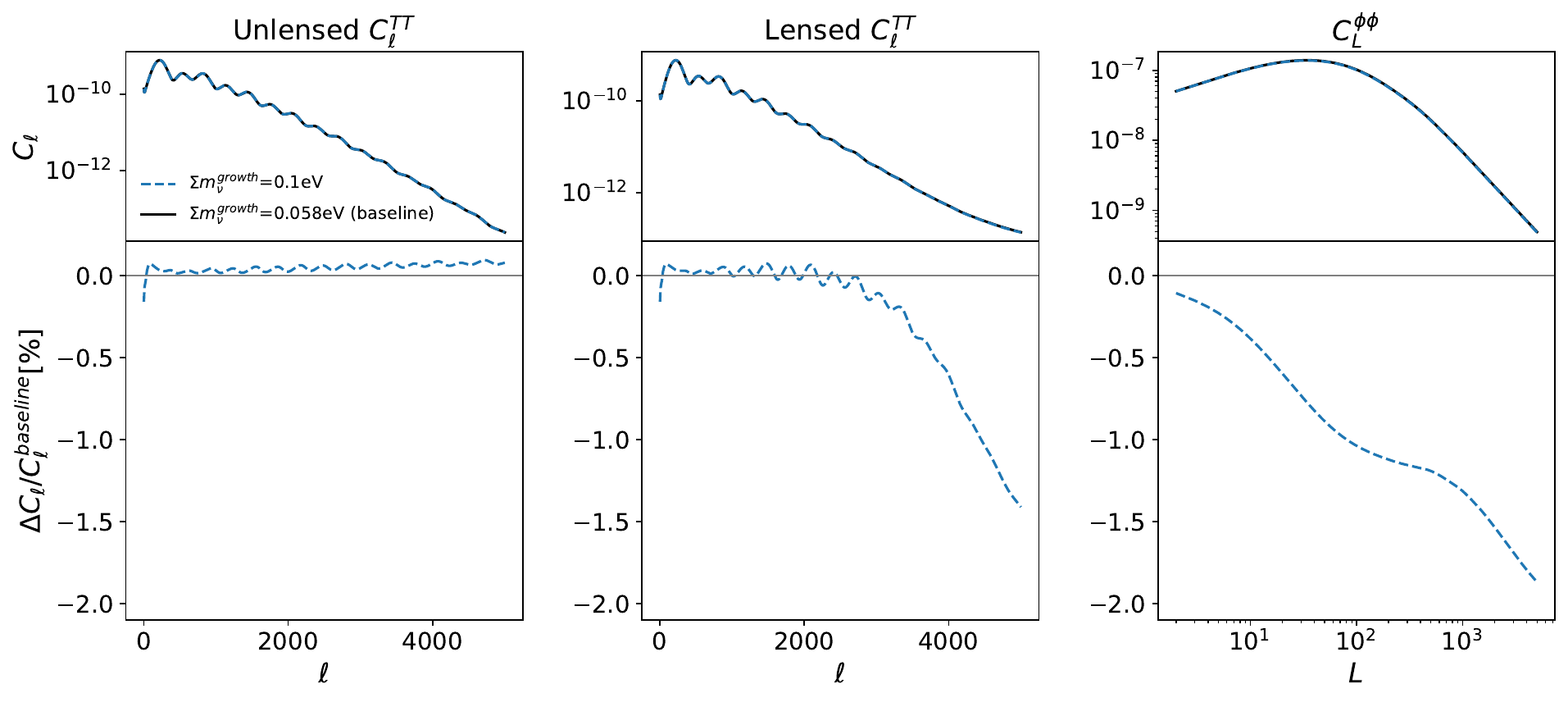}
    \caption{Similar to Fig.~\ref{fig:geo_effect}, but here shows the effects on CMB temperature and lensing power spectra when changing $\mgrowth$ with holding $\mgeo=0.058~\mathrm{eV}$ . The unlensed CMB TT spectrum is in the left panels, and the lensed one is in the middle. The top panels show the power spectra; the bottom panels show the fractional difference of the prediction for $\mgrowth = 0.1$eV compared to the baseline $\mgrowth = 0.058$eV result. On the right shows the plots for the CMB lensing potential power spectrum, $C_L^{\phi\phi}$.
    }
    \label{fig:growth_effect}
\end{figure*}

In Fig.~\ref{fig:growth_effect}, we show the changes to the CMB temperature and lensing power spectra induced by varying $\mgrowth$. The perturbation evolution differs for different $\mgrowth$ values, while the geometrical projection effects are the same among these models since we fix $\mgeo=0.058~\mathrm{eV}$ in the figure.

The most obvious changes induced by $\mgrowth$ are in the CMB lensing potential, compared to those induced by $\mgeo$. Increasing $\mgrowth$ suppresses structure growth through neutrino free-streaming, which in turn reduces the amplitude of the CMB lensing potential power spectrum. The most significant changes to $C_L^{\phi\phi}$ occur at small angular scales, with approximately 2\% decrease in lensing power spectrum at $L\sim 4000$. It is expected since free-streaming massive neutrinos suppress the power spectrum more strongly at smaller scales.

On the contrary, the changes to the unlensed $C_\ell^{TT}$ are quite small compared to the changes in lensing. The fractional changes of unlensed $C_\ell^{TT}$ is within $0.2\%$ throughout $100\lesssim\ell$. This is caused by the integrated Sachs-Wolfe (ISW) effect at late time as well as the changes in the diffusion damping scale in response to the change in $\mgrowth$.

Given the response of the unlensed $C_\ell^{TT}$ and $C_L^{\phi\phi}$ to the change in $\mgrowth$, the lensed CMB temperature spectrum also shows minimal changes at $\ell\lesssim2000$ and more significant changes at small angular scales with approximately $1.5\%$ decrease in power at $\ell \sim 4600$. 
Since the small-scale unlensed CMB temperature power is damped and the lensing effect contributes significantly to the power of the lensed CMB at those scales, the changes to the lensed $C_\ell^{TT}$ are expected. 
At large scales, where the CMB lensing effect is less significant, the fractional changes in the lensed $C_\ell^{TT}$ are similar to those in the unlensed $C_\ell^{TT}$.

\section{Results}
\label{sec:split_results}

After defining the $\mgeo$ and $\mgrowth$ split implementation and seeing their impacts on BAO and CMB observables, we now turn to the constraints on $\mgeo$ and $\mgrowth$. As discussed in Sect.~\ref{sec:fiducial_swap}, the truncated-Gaussian posterior conflates a genuine tightening of the constraint with a shift of the underlying peak. Thus, we first show constraints from a Fisher forecast and from MCMC runs on mock datasets in Sect.~\ref{sec:result_fisher}, both at the measurement precision of current observations. We then present results with real data in Sect.~\ref{sec:result_real}, both under the standard non-negative prior and with the parameter range extrapolated to negative values.

To explore possible preferences for the opposite of massive neutrinos, we extend to 
``negative'' values of $\mgeo$ and $\mgrowth$, following the method used in \citep{Elbers:2024sha, elbers2025constraintsneutrinophysicsdesi, Sailer:2025lxj}. Since the neutrino mass sum is close to zero and the effect of non-zero $\sum m_\nu$ on CMB and BAO observables is small, \citep{Elbers:2024sha} proposes to extrapolate $\sum m_\nu$ to negative values at the linear order via
\begin{equation}
    X(\sum\tilde{m}_\nu) = X_0 + {\rm Sgn}(\sum\tilde{m}_\nu)\left[X(|\sum \tilde{m}_\nu|) - X_0\right],
\end{equation}
where $X$ represents the cosmological observables, $X_0 = X(\sum m_\nu = 0)$, and ${\rm Sgn}(\sum\tilde{m}_\nu)$ is defined in Eqn.~(\ref{eqn:sign}).
In our case, where $\mnu$ is divided into $\mgeo$ and $\mgrowth$, we apply the same linear extrapolation when either $\mgeo<0$ or $\mgrowth<0$
\begin{align}\label{eqn:negative}
    &X(\sum{m}_\nu^{\rm geo}, \sum{m}_\nu^{\rm growth}) \nonumber \\
    = &X_0 + {\rm Sgn}(\sum{m}_\nu^{\rm geo})\left[X(|\sum{m}_\nu^{\rm geo}|, 0) - X_0\right] \nonumber \\
    &+ {\rm Sgn}(\sum{m}_\nu^{\rm growth})\left[X(0, |\sum{m}_\nu^{\rm growth}|) - X_0\right],
\end{align}
where $X_0 = X(\sum{m}_\nu^{\rm geo} = 0, \sum{m}_\nu^{\rm growth} = 0)$. To implement this, we modified the publicly available code\footnote{\url{https://github.com/NoahSailer/disputauble}} from \citep{Sailer:2025lxj}.

To get the posterior distribution of $\mgeo$ and $\mgrowth$, we use \texttt{Cobaya}\footnote{\url{https://github.com/CobayaSampler/cobaya}} sampler \citep{Torrado:2020dgo} for running Markov Chain Monte Carlo for the results. 

\subsection{Geometry Channel is Slightly More Constraining than Growth Channel} \label{sec:result_fisher}

In this section, we discuss the constraints on $\mgeo \ \& \ \mgrowth$ from a Fisher forecast with measurement precisions of the current data, and from Monte Carlo Markov Chain runs on mock data. 
Working with forecasts and mock data lets us separate the statistical constraining power of each data combination from the shift of the posterior caused by the mild tension between different datasets, which is hard to disentangle for $\sum m_\nu$ whose prior truncates at zero.

\begin{figure}
    \centering
    \includegraphics[width=0.95\linewidth]{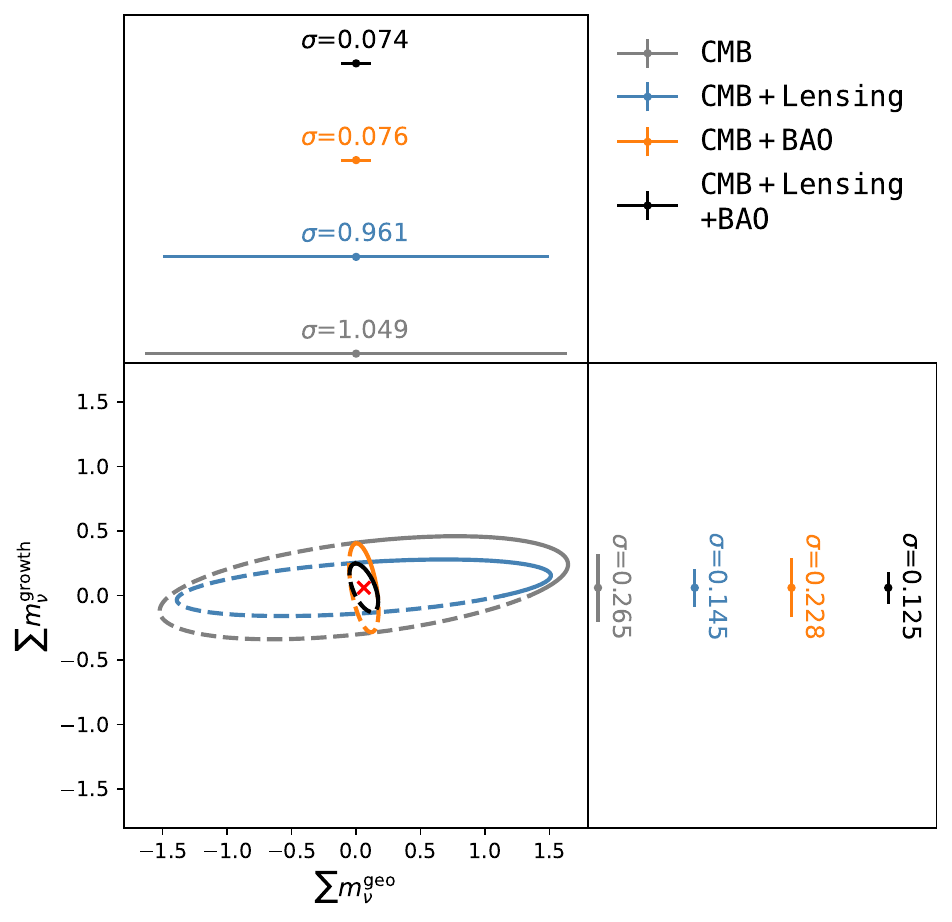}
    \caption{Forecasted 2D posteriors of $\mgeo \ \& \ \mgrowth$ with \texttt{CamSpec} and \texttt{lensing} and \texttt{BAO} likelihoods. The uncertainties shown in the plot are derived from Fisher forecast using the real covariance of each measurement. The contours and errorbars are quoted for 68\% C.L. The dashed line segments represent the case when either $\mgeo$ or $\mgrowth$ becomes negative; the solid line segments show when both $\mgeo$ and $\mgrowth$ are non-negative. The red cross represents the fiducial value with $\mgeo=\mgrowth=0.058$~eV.
    Both geometry and growth channels are similarly constrained. The geometry channel, mostly informed by CMB and BAO, is slightly better constrained than the growth channel, mostly informed by CMB lensing.
    }
    \label{fig:mnu_fisher_camspec_geo_growth}
\end{figure}

The forecast constraints are shown in Fig.~\ref{fig:mnu_fisher_camspec_geo_growth} and collected in Table~\ref{tab:fiducial}. From the \texttt{CMB} alone, and from the \texttt{CMB+Lensing}, $\mgrowth$ is the better measured of the two parameters; as soon as \texttt{BAO} is included, the ordering reverses and $\mgeo$ becomes the better measured one. Combining all three probes,
\begin{align}\label{eq:fisher_all}
    \sigma(\mgeo) &= 0.074~{\rm eV}, \\ \nonumber
    \sigma(\mgrowth) &= 0.125~{\rm eV};
\end{align}
the neutrino mass sum inferred from the geometry channel is a factor of $1.7$ better measured than the one inferred from the growth channel. The two channels are limited by different physics.

From primary CMB alone, neither parameter is tightly constrained with $\sigma(\mgeo) = 1.049~{\rm eV}$ against $\sigma(\mgrowth) = 0.265~{\rm eV}$, shown in gray in Fig.~\ref{fig:mnu_fisher_camspec_geo_growth}. 
As discussed in Sect.~\ref{sec:mnu_geo_cl}, $\mgeo$ enters the primary CMB only through the late-time expansion rate, and therefore via $D_A(z_*)$. With well-measured CMB peaks and troughs, the $\theta_*$ is tightly constrained. The degeneracy between $\omega_c-H_0-\mgeo$ emerges in the integral of $D_A(z_*)$, where
an increase in $\mgeo$ is thus absorbed by a lower $H_0$ together with a small adjustment of $\omega_c$. Thus, the primary CMB on its own has no handle with which to separate the three. 
A forecast uncertainty of order $1~{\rm eV}$ should therefore be read as ``unconstrained'' rather than as a measurement.
$\mgrowth$, by contrast, is measured even without the lensing power spectrum, because the primary CMB spectra we use are \textit{lensed}: the peak smearing and the transport of power to small scales described in Sect.~\ref{sec:mnu_growth_cl} already constitute a measurement of the lensing amplitude, and hence of the growth channel. 
We have verified that this is the origin of the difference by repeating the forecast with unlensed spectra, in which case $\sigma(\mgrowth) = 2.3~{\rm eV}$ and $\sigma(\mgeo) = 3.0~{\rm eV}$: the growth advantage disappears and the constraints from the two channels become comparable, as expected if the sensitivity of the primary CMB to $\mgrowth$ is carried entirely by lensing.

Including the \texttt{Lensing} likelihood, shown in blue, tightens $\sigma(\mgrowth)$ from $0.265$ to $0.145~{\rm eV}$, an improvement of $45\%$, as CMB lensing is a probe of structure growth. While $\mgeo$ improves by only $8\%$, from $1.049$ to $0.961~{\rm eV}$, a reduction merely due to $\sigma(\mgrowth)$ improvement along the $\mgeo-\mgrowth$ degeneracy direction. 

Adding \texttt{BAO} to \texttt{CMB} tightens $\sigma(\mgeo)$ from $1.049$ to $0.076~{\rm eV}$, a factor of $14$ improvement. This is because BAO pins $\Omega_m$ independently of the CMB and so fixes the late-time expansion history that $\mgeo$ controls. 
The growth channel also improves, from $0.265$ to $0.228~{\rm eV}$, and this improvement is necessarily indirect. By construction, the BAO observables in our model depend only on the geometry sector (Fig.~\ref{fig:flowchart}), so they carry no information about $\mgrowth$ of their own, and the gain comes entirely from the $\mgeo$--$\mgrowth$ degeneracy. 

\begin{table}[ht]
\centering
\small
\setlength{\tabcolsep}{4pt}
\begin{tabular}{l | c c | c c}
\hline
 & \multicolumn{2}{c|}{$\mgeo$ [eV]} & \multicolumn{2}{c}{$\mgrowth$ [eV]} \\
Data & MCMC & $\sigma_{\rm Fisher}$ & MCMC & $\sigma_{\rm Fisher}$ \\
\hline
\texttt{CMB}\footnotemark[1] & $-0.24^{+0.55}_{-0.81}$ & $1.049$ & $-0.38^{+0.41}_{-0.31}$ & $0.265$ \\
\texttt{CMB+L(mock)} & $0.08^{+0.47}_{-0.60}$ & $0.961$ & $-0.03\pm 0.15$ & $0.145$ \\
\texttt{CMB+B(mock)} & $0.12\pm 0.11$ & $0.076$ & $-0.23^{+0.28}_{-0.21}$ & $0.228$ \\
\texttt{CMB+B+L(mock)} & $0.113\pm{0.095}$ & $0.074$ & $-0.03\pm 0.11$ & $0.125$ \\
\hline
\end{tabular}
\footnotetext[1]{The \texttt{CMB} row uses the default \texttt{CamSpec} likelihood; only the \texttt{Lensing} and \texttt{BAO} data vectors are replaced by mock predictions.}
\caption{Constraints on $\mgeo$ and $\mgrowth$ with mock CMB lensing and BAO data, alongside the Fisher uncertainties of Fig.~\ref{fig:mnu_fisher_camspec_geo_growth}. The MCMC entries are $68\%$ intervals, with the parameter range extrapolated to negative values following Eqn.~(\ref{eqn:negative}). We use the shorthand \texttt{L} for \texttt{Lensing} and \texttt{B} for \texttt{BAO} as in Fig.~\ref{fig:fiducial}.}
\label{tab:fiducial}
\end{table}

We now turn to the mock-data runs. We substitute the data vectors in the \texttt{BAO} and \texttt{Lensing} likelihoods with the predictions of the \texttt{CamSpec} best-fit cosmology, which has $\mgeo = \mgrowth = 0.058$~eV, and we open up the parameter space to negative $\mgeo$ and $\mgrowth$ following Eqn.~(\ref{eqn:negative}). 
Together these remove the two effects that complicate the interpretation of the real-data posteriors, namely the shift caused by the excess lensing power and by the mild tension between CMB and BAO, and the truncation of the posterior at zero. 
For \texttt{CMB} we keep the default \texttt{CamSpec} likelihood without substituting the data vector. 
Note that this choice of mock cosmology, unlike the joint \texttt{CMB+BAO} best fit also used in Sect.~\ref{sec:fiducial_swap}, treats the CMB best fit as the truth; the central values below should be read with that in mind. 
The results are 
summarized in Table~\ref{tab:fiducial}.

The mock runs reproduce the pattern of the forecast: the constraints from \texttt{CMB} alone are very weak, adding CMB lensing improves $\mgrowth$ substantially, and adding BAO improves $\mgeo$ by more still. 
The tightest constraints come from \texttt{CMB+Lensing+BAO}, which yields
\begin{align}\label{eq:mock_all}
    \sigma(\mgeo) &= 0.095~{\rm eV}, \\ \nonumber
    \sigma(\mgrowth) &= 0.11~{\rm eV}.
\end{align}
Similarly to the forecast, $\mgeo$ is the better constrained of the two. 
Comparing Eqn.~(\ref{eq:mock_all}) to Eqn.~(\ref{eq:fisher_all}), the Fisher forecast gives constraints on $\mgeo$ and $\mgrowth$ more optimistically by $28\%$ and $12\%$ than MCMC results. Thus, we will discuss the $\mgeo$ and $\mgrowth$ constraints from MCMC in the Sect.~\ref{sec:result_real}.

\subsection{Geometry and growth channel both slightly favor negative values} \label{sec:result_real}

We now turn to the real data. 
Replacing the mock \texttt{Lensing} and \texttt{BAO} data vectors of Sect.~\ref{sec:result_fisher} with the real ones leaves the uncertainties on both parameters essentially unchanged but moves their central values downward, and about twice as far in the geometry channel as in the growth channel (see Table~\ref{tab:fiducial}\&\ref{tab:disputauble}).
Although the geometry and growth channels show preference for a neutrino mass sum lower than $0.058~\mathrm{eV}$ at approximately $1.3~\sigma$ and $1.8~\sigma$ level respectively, we also find that neither channel on its own excludes the minimum mass sum allowed by oscillation experiments. 
In the following, we first show the constraints with the range extended to the negative regime, then the canonical non-negative results.

\begin{figure}[!]
    \centering
    \includegraphics[width=0.95\linewidth]{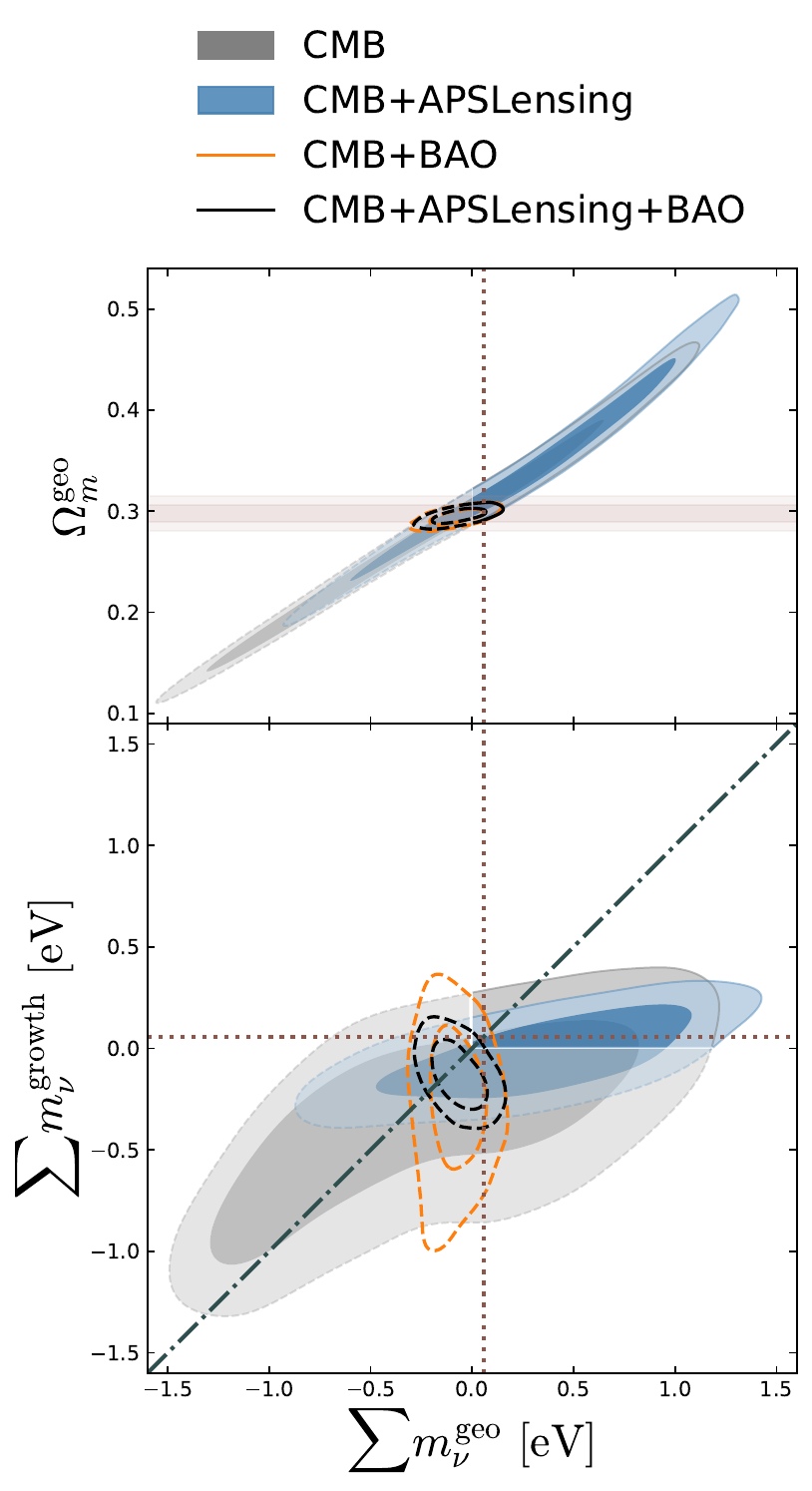}
    \caption{Marginal 2D posteriors of $\mgrowth \ vs \ \mgeo$ in the bottom panel and $\Omega^\mathrm{geo}_m \ vs \ \mgeo$ in the top panel ($\Omega^\mathrm{geo}_m = \Omega_b + \Omega_c + \Omega_\nu^\mathrm{geo}$), on real data. 
    We show the region where either $\mgeo$ or $\mgrowth$ is negative in lighter filled regoins or dashed lines. The vertical and horizontal dotted lines indicate the neutrino mass sum at 0.058 eV. The dash-dotted line in the bottom panel indicates $\mgeo = \mgrowth$. The horizontal band in the top panel shows the $\Omega_m$ constraints from \texttt{BAO}. Combining \texttt{CMB}+\texttt{Lensing}+\texttt{BAO}, both geometry and growth channel slightly prefer the ``negative neutrino'' mass. The fiducial NO mass of $0.058~\mathrm{eV}$ is excluded out side of the 95\% C.L. region. The difference $\Delta \sum m_\nu = \mgeo - \mgrowth$ remains consistent with zero, passing the internal consistency test of $\Lambda$CDM.
    }
    \label{fig:mnu_only_geo_growth}
\end{figure}

In Table~\ref{tab:disputauble} and Fig.~\ref{fig:mnu_only_geo_growth}, we show the constraints of $\mgeo \ \& \ \mgrowth$ when they are extrapolated to negative domain following Eqn.~(\ref{eqn:negative}). 
Using \texttt{CMB} alone, the uncertainties on both $\mgeo$ and $\mgrowth$ are quite large, with $\mgrowth$ slightly tighter, qualitatively similar to the behavior of the $\mgeo$ and $\mgrowth$ shown in the Sect.~\ref{sec:result_fisher}. 
Adding \texttt{Lensing} tightens $\mgrowth$ to $-0.04\pm0.15$~eV, since the CMB lensing power spectrum is sensitive to the perturbation evolution through $\mgrowth$ as discussed in Sect.~\ref{sec:mnu_growth_cl}. 
It also moves the mean of $\mgeo$ upward from $-0.24$ to $0.33$~eV, without improving its width appreciably;
this is sourced by adding \texttt{lensing} to \texttt{CMB} where $\mgrowth$ tightens and selects the upper right volume of the \texttt{CMB}-only posterior (the gray contour in Fig.~\ref{fig:mnu_only_geo_growth}).

\begin{table}[ht]
\centering
\begin{tabular}{l | c  c}
\hline
Data & $\mgeo$ [eV] & $\mgrowth$ [eV] \\
\hline
\texttt{CMB} & $-0.24^{+0.55}_{-0.81}$ & $-0.38^{+0.41}_{-0.31}$ \\
\texttt{CMB+Lensing} & $0.33^{+0.58}_{-0.40}$ & $-0.04\pm 0.15$ \\
\texttt{CMB+BAO} & $-0.090^{+0.099}_{-0.081}$ & $-0.26^{+0.24}_{-0.21}$ \\
\texttt{CMB+BAO+Lensing} & $-0.07\pm 0.10$ & $-0.13^{+0.10}_{-0.12}$ \\
\hline
\end{tabular}
\caption{Constraints on $\mgeo \ \& \ \mgrowth$ when extrapolating those parameter space to negative values. All limits quoted are 68\% C.L.}
\label{tab:disputauble}
\end{table}

Adding \texttt{BAO} to \texttt{CMB} instead improves $\mgeo$ directly, since BAO constrains the background geometry and the impact of $\mgeo$ is confined to geometry channel only. The uncertainty on $\mgrowth$ also decreases, through degeneracy breaking, as shown by the orange contours of Fig.~\ref{fig:mnu_only_geo_growth}.

It is also interesting to notice that the improvement on $\sigma(\mgrowth)$ by adding \texttt{BAO} to \texttt{CMB} through this degeneracy breaking is more significant than that on $\sigma(\mgeo)$ by adding \texttt{Lensing} to \texttt{CMB}, which is also seen in the mock MCMC runs of Sect.~\ref{sec:result_fisher} and Table~\ref{tab:fiducial}.
The top panel of Fig.~\ref{fig:mnu_only_geo_growth} shows the reason. In the split model there is no longer a single matter density: the geometry sector has its own $\Omega_m^{\rm geo} = \Omega_b + \Omega_c + \Omega_\nu^{\rm geo}$, which is what BAO measures. Because the primary CMB fixes $\omega_b$ and $\omega_c$ tightly, $\Omega_m^{\rm geo}$ and $\mgeo$ are confined to a narrow line in the plane but are individually poorly determined along it. The tight constraint from \texttt{BAO} on $\Omega_m^{geo}$ is shown as the narrow horizontal band, which cuts that line and localizes $\mgeo$, helping to improve the constraint on $\mgrowth$.

Combining all \texttt{CMB}, \texttt{Lensing} and \texttt{BAO}, we get
\begin{align}\label{eq:real_all}
    \mgeo &= -0.07\pm0.10~{\rm eV}, \\ \nonumber
    \mgrowth &= -0.13^{+0.10}_{-0.12}~{\rm eV}.
\end{align}
We note that the mock runs of Sect.~\ref{sec:result_fisher} and the runs above differ only in the \texttt{Lensing} and \texttt{BAO} data vectors, so comparing them isolates what those data do to the split. 
Between Table~\ref{tab:fiducial} and Table~\ref{tab:disputauble}, the uncertainties barely move, $\sigma(\mgeo) \simeq 0.10$~eV and $\sigma(\mgrowth)$ remains around  $0.11$~eV, while the central values move from $\mgeo = 0.113$ to $-0.07$~eV and from $\mgrowth = -0.03$ to $-0.13$~eV, that is by roughly $1.9\sigma$ and $0.9\sigma$ respectively.
The real data therefore shift the posterior,
and more of the shift is carried by the geometry channel under the assumption that the true cosmology is the one preferred by \texttt{CMB}, because the mock data in Sect.~\ref{sec:result_fisher} is based on \texttt{CMB} best-fit cosmology. 
This is the conclusion of Sect.~\ref{sec:fiducial_swap}, that the mild $\Omega_m$ disagreement between CMB and BAO is what the neutrino mass sum is forced to absorb, now expressed in terms of the split. The two runs share the same real \texttt{CamSpec} likelihood, so the two shifts are correlated and the quoted significances should be read as indicative. 

Our split differs from the background-only approach of \citep{lynch2025whatsmattersigmamnu} (see Sect.~\ref{sec:review}) in which the extension to negative values is applied. 
They reverse the sign of the neutrino rest-mass contribution to the energy density, which leaves the perturbations undefined and forces the primary CMB to be compressed into a Gaussian in $(\omega_b, \omega_c, \theta_s^*)$; the growth sector is then an assumption made in building that compression rather than a fitted quantity, and their result moves by $0.106$~eV, which is similar to its own uncertainty, according to whether the compressed likelihood is built from the full $\Lambda$CDM posterior distribution or $\Lambda$CDM+$A_{\rm 2pt}$ posterior distribution. 
Extrapolating the observables instead, with Eqn.~(\ref{eqn:negative}), keeps the perturbation evolution well defined and the full CMB likelihood usable. Our $\mgeo$ is consistent with all their variants while less negative than any, disfavouring $0.058$~eV at $1.3\sigma$. It is expected, because a deficit that their single parameter must absorb alone is here shared with $\mgrowth$, as well as slightly difference in data choice. 

The same distinction applies to \citep{graham2025darkforcesgathering}, whose $\sum m_\nu^{\rm clustering}=-0.15\pm0.16$ and $\sum m_\nu^{\rm BAO}=-0.032\pm0.099$ agree with our $\mgrowth$ and $\mgeo$, but they are defined through individual observables rather than through the physical effects from geometry and growth. The data sets used for deriving the constraints are also different: the \texttt{CMB} consists of \texttt{CamSpec} likelihood only and \texttt{lensing} is the joint lensing likelihood of ACT, \textit{Planck} and SPT-3G; both the primary CMB and CMB lensing likelihoods are combinations of \textit{Planck} and ACT-DR6 in \citep{graham2025darkforcesgathering} following \citep{act_2}.

Note that we used the cosmological data across the cosmic history from last scattering until today to get the constraints in Eqn.~(\ref{eq:real_all}). One would expect a different result if following the data cut of \citep{Qu:2026fkt} to use a late-Universe-free data. We leave constraining $\mgeo$ and $\mgrowth$ using the late-Universe-free data for a future work.

Since the $\Lambda$CDM assume $\mgeo=\mgrowth$, we also report the difference of $\sum m_\nu$ from the geometry and growth channel, defined as $\mdiff \equiv \mgeo - \mgrowth$. The full \texttt{CMB+Lensing+BAO} combination gives
\begin{equation}\label{eq:delta}
    \mdiff = 0.07\pm 0.17~{\rm eV},
\end{equation}
consistent with zero at $\sim0.4\sigma$ level. The predictions of the background geometry and of the perturbation evolution therefore agree within the present sensitivity, and the $\Lambda$CDM internal consistency test is passed. We note that this is a statement about the \textit{difference}: both parameters are pulled low together, and the minimum mass sum of the normal ordering, $\mgeo = \mgrowth = 0.058$~eV, lies outside the $2\sigma$ region of the joint posterior. In other words, current data are consistent with geometry and growth telling the same story, and that shared story is in mild tension with the oscillation minimum. 
Our findings are in agreement with \citep{graham2025darkforcesgathering}.

\begin{figure}
    \centering
    \includegraphics[width=0.95\linewidth]{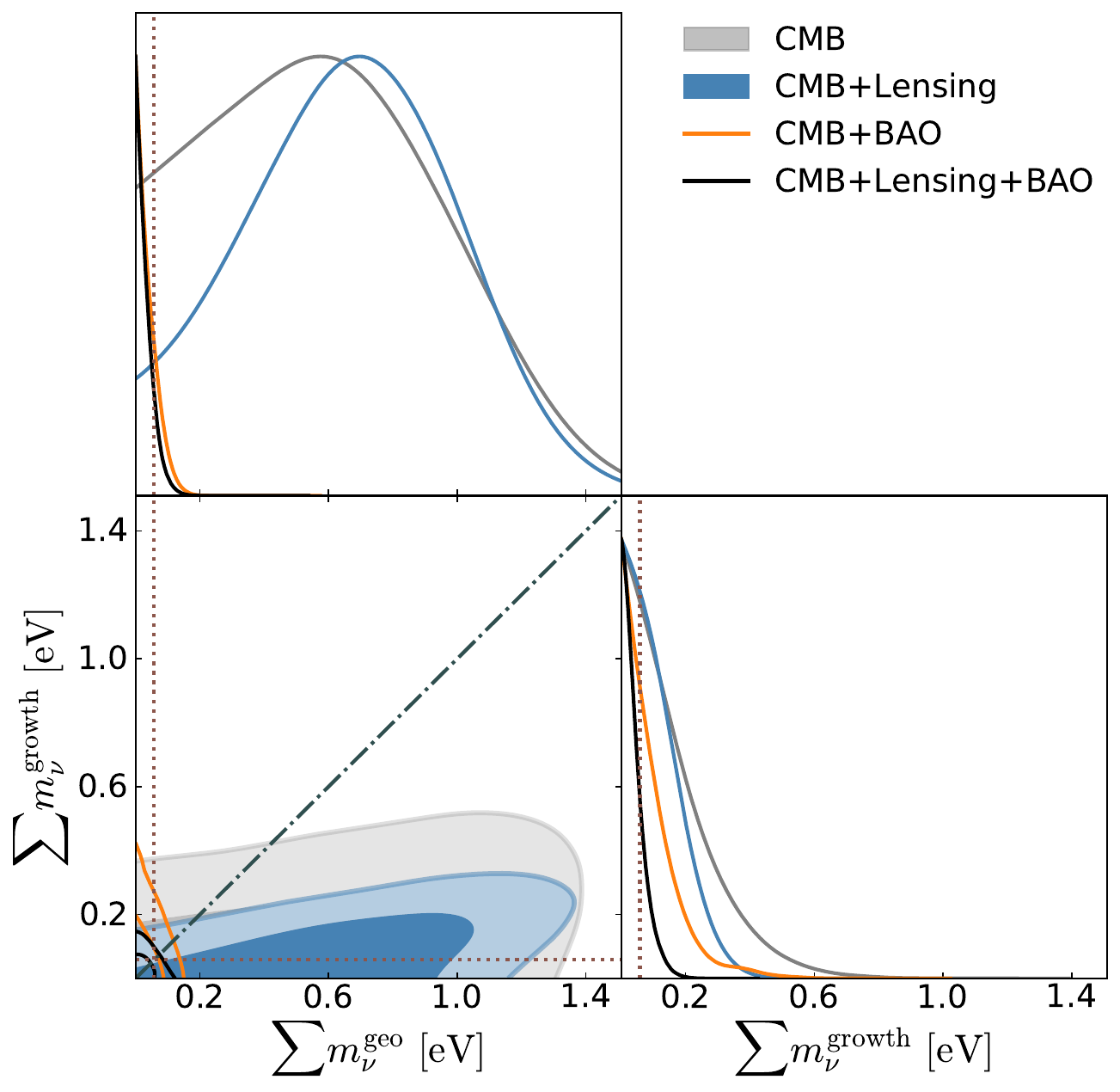}
    \caption{Marginalized posterior on positive $\mgeo$ and $\mgrowth$ from real data. The dotted lines show the minimum neutrino mass sum at normal hierarchy at 0.058 eV. The dash-dotted line is for $\mgeo=\mgrowth$ case.}
    \label{fig:mnu_geo_growth}
\end{figure}

Finally, in Fig.~\ref{fig:mnu_geo_growth} and Table~\ref{tab:nonnegative} we show the constraints obtained under the canonical prior that both parameters be non-negative. The pattern is the one established above: both are loosely constrained by \texttt{CMB} alone, with $\mgrowth$ the tighter, and adding \texttt{Lensing} or \texttt{BAO} improves $\mgrowth$ or $\mgeo$ respectively. Combining \texttt{CMB}, \texttt{Lensing} and \texttt{BAO} yields
\begin{align}\label{eq:nonneg_all}
    \mgeo &< 0.081~{\rm eV}, \\ \nonumber
    \mgrowth &< 0.11~{\rm eV},
\end{align}
at $95\%$ C.L.

\begin{table}[ht]
\centering
\begin{tabular}{l | c  c}
\hline
Data & $\mgeo$ [eV] & $\mgrowth$ [eV] \\
\hline
\texttt{CMB} & $< 1.22$ & $< 0.43$ \\
\texttt{CMB+Lensing} & $<1.21$ & $< 0.27$ \\
\texttt{CMB+BAO} & $< 0.102$ & $< 0.262$ \\
\texttt{CMB+BAO+Lensing} & $< 0.0814$ & $< 0.113$ \\
\hline
\end{tabular}
\caption{Constraints on non-negative $\mgeo \ \& \ \mgrowth$ with priors that both $\mgeo \ \& \ \mgrowth$ are non-negative. All limits quoted are 95\% C.L.}
\label{tab:nonnegative}
\end{table}

Both of these bounds lie above the minimum mass sum of the NO at $0.058$~eV, whereas the single-parameter analysis of the same data gives $\sum m_\nu < 0.0597$~eV, Eqn.~(\ref{eq:mnu2}). 
Neither the geometrical information nor the perturbative information in current CMB and BAO data excludes the oscillation minimum on its own; the tighter 95\% C.L. upper bound for neutrino mass sum appears only once the two are locked together by the assumption $\mgeo = \mgrowth$.

\begin{figure}
    \centering
    \includegraphics[width=0.95\linewidth]{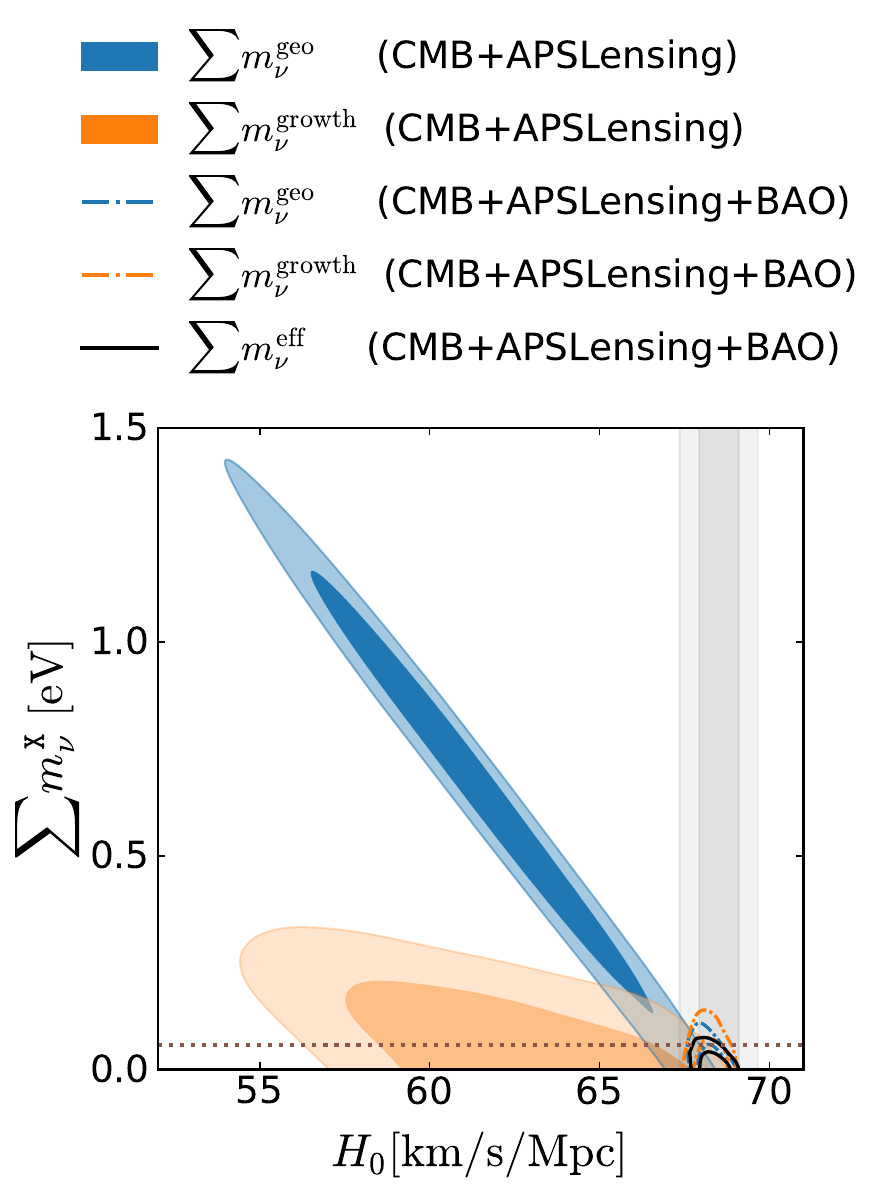}
    \caption{Marginal posteriors of $\Sigma m_\nu^\texttt{X}$ v.s. $H_0$, where superscript $\texttt{X}$ represents \textit{geo}, \textit{growth} of our split implementation or the single effective neutrino mass sum by setting $\meff=\mgeo=\mgrowth$. The horizontal dotted line shows the minimal neutrino mass sum at 0.058 eV. The gray vertical band is quoted from DESI DR2 result for illustration.
    The degeneracy directions of $\mgeo$ and $\mgrowth$ with $H_0$ are different. The tight constraints on single parametrization of $\meff=\mgeo=\mgrowth$ are due to the degeneracy breaking.}
    \label{fig:mnu_geo_growth_breaking}
\end{figure}

Fig.~\ref{fig:mnu_geo_growth_breaking} shows how that happens, by plotting $\Sigma m_\nu^\texttt{X}$ against $H_0$, where the superscript $\texttt{X}$ denotes ``$geo$'' or ``$growth$'' for our split and ``$eff$'' for the case in which we impose $\meff = \mgeo = \mgrowth$. The two channels are degenerate with $H_0$ in different directions. The blue $\mgeo$ contour from \texttt{CMB+Lensing} runs steeply down in $H_0$: this is the fixed-$\theta_*$ trade-off anticipated in Sect.~\ref{sec:result_fisher}, now seen in the posterior, and it is what limits the primary-CMB constraint on the geometry channel. The orange $\mgrowth$ contour is much flatter. Adding \texttt{BAO}, which supplies an $H_0$ constraint (gray vertical band), tightens each to the dashed regions, and imposing $\meff = \mgeo = \mgrowth$ selects the intersection of the two, giving the black contour. The tight upper bound on the neutrino mass sum in the standard analysis is thus a product of combining two channels with different degeneracy directions, and its strength depends on the assumption that the same mass parameter governs both.

\section{Forecasts: the two channels as a diagnostic of model error}\label{sec:forecast}

In this section, we forecast the constraints on $\mgeo$ and $\mgrowth$ for future CMB and BAO data, following the procedure and setup of \citep{LSSTDarkEnergyScience:2026ach}. For the CMB we consider two upcoming surveys, the South Pole Telescope (SPT-3G) \citep{SPT-3G:2024qkd, SPT-3G:2025mju} and the advanced Simons Observatory (ASO) \citep{SimonsObservatory:2025wwn}, together with a futuristic ground-based survey matching the noise and angular resolution of CMB-S4 \citep{CMB-S4:2016ple, Abazajian:2019eic}, combining TT, TE, EE and $\phi\phi$ power spectra in each case. For BAO we use the forecast for the DESI DR3 release \citep{DESI:2023dwi}. We use and modify the likelihood and the Fisher forecast code \texttt{CMB\_BAO\_SNe\_likelihoods}\footnote{\url{https://github.com/sriniraghunathan/CMB_BAO_SNe_likelihoods}}, and for the Boltzmann solver we use the modified \texttt{CAMB} \citep{Lewis:1999bs, Howlett:2012mh} of Sect.~\ref{sec:split_imp}. All forecasts are evaluated at a fiducial $\mgeo = \mgrowth = 0.058$~eV.

Table~\ref{tab:forecast} summarizes the results.
Without BAO, $\mgrowth$ is the better measured parameter for every CMB configuration, for the reason given in Sect.~\ref{sec:result_fisher}, and it improves steadily with CMB sensitivity, from $0.14$~eV for SPT-3G to $0.087$~eV for CMB-S4. The geometry channel improves over the same range as well, from $0.78$ to $0.30$~eV, but remains an order of magnitude weaker. Once DESI DR3 BAO is added the picture inverts: $\sigma(\mgeo)$ collapses to $0.050$, $0.043$ and $0.038$~eV for the three surveys, a factor of $15$ to $8$ improvement, and the three configurations now differ by only about $32\%$, whereas without BAO they span a factor of $2.6$. The geometry channel is therefore essentially BAO-limited once DESI DR3 is available, while the growth channel continues to be set by the CMB, improving from $0.097$ to $0.072$~eV across the same three surveys. Future gains in the two channels will come from different experiments.

Comparing with the current constraints of Table~\ref{tab:disputauble}, CMB-S4 alone would improve on the present constraint of  $\mgrowth$ from \texttt{CMB+Lensing+BAO} without using any BAO information at all, while ASO Baseline would match it. Adding DESI DR3 BAO brings both channels below $0.10$~eV for every configuration considered.

Although the $\mgeo$ uncertainty from Fisher forecast is shown to be about $28\%$ more optimistic than that derived from MCMC analysis (see Sect.~\ref{sec:result_fisher}), we expect the result presented here to be more accurate with the improved constraining power from the data.

If the negative central value of $\mgeo = -0.07$~eV found in Eqn.~(\ref{eq:real_all}) persists, ASO Baseline and CMB-S4 combined with DESI DR3 BAO would place the minimum mass sum of $0.058$~eV at $3.0\sigma$ and $3.4\sigma$ from the measured value, using the geometrical information alone.

\begin{table}[ht]
\centering
Without DESI DR3 BAO
\begin{tabular}{l | c  c}
\hline
CMB Data Set & $\sigma(\mgeo)$ [eV] & $\sigma(\mgrowth)$ [eV] \\
\hline
\texttt{SPT-3G (Ext-10k)} & $0.78$ & $0.14$ \\
\texttt{ASO Baseline} & $0.59$ & $0.11$ \\
\texttt{CMB-S4} & $0.30$ & $0.087$ \\
\hline
\end{tabular}\\
\vspace{5pt}
With DESI DR3 BAO
\begin{tabular}{l | c  c}
\hline
CMB Data Set & $\sigma(\mgeo)$ [eV] & $\sigma(\mgrowth)$ [eV] \\
\hline
\texttt{SPT-3G (Ext-10k)} & $0.050$ & $0.097$ \\
\texttt{ASO Baseline} & $0.043$ & $0.082$ \\
\texttt{CMB-S4} & $0.038$ & $0.072$ \\
\hline
\end{tabular}
\caption{Forecasted uncertainty of $\mgeo \ \& \ \mgrowth$. We use TT, TE, EE and $\phi\phi$ power spectra for the CMB data. The top and bottom tables show the forecast uncertainties without and with DESI DR3 BAO respectively.}
\label{tab:forecast}
\end{table}

\begin{figure}
    \centering
    \includegraphics[width=0.95\linewidth]{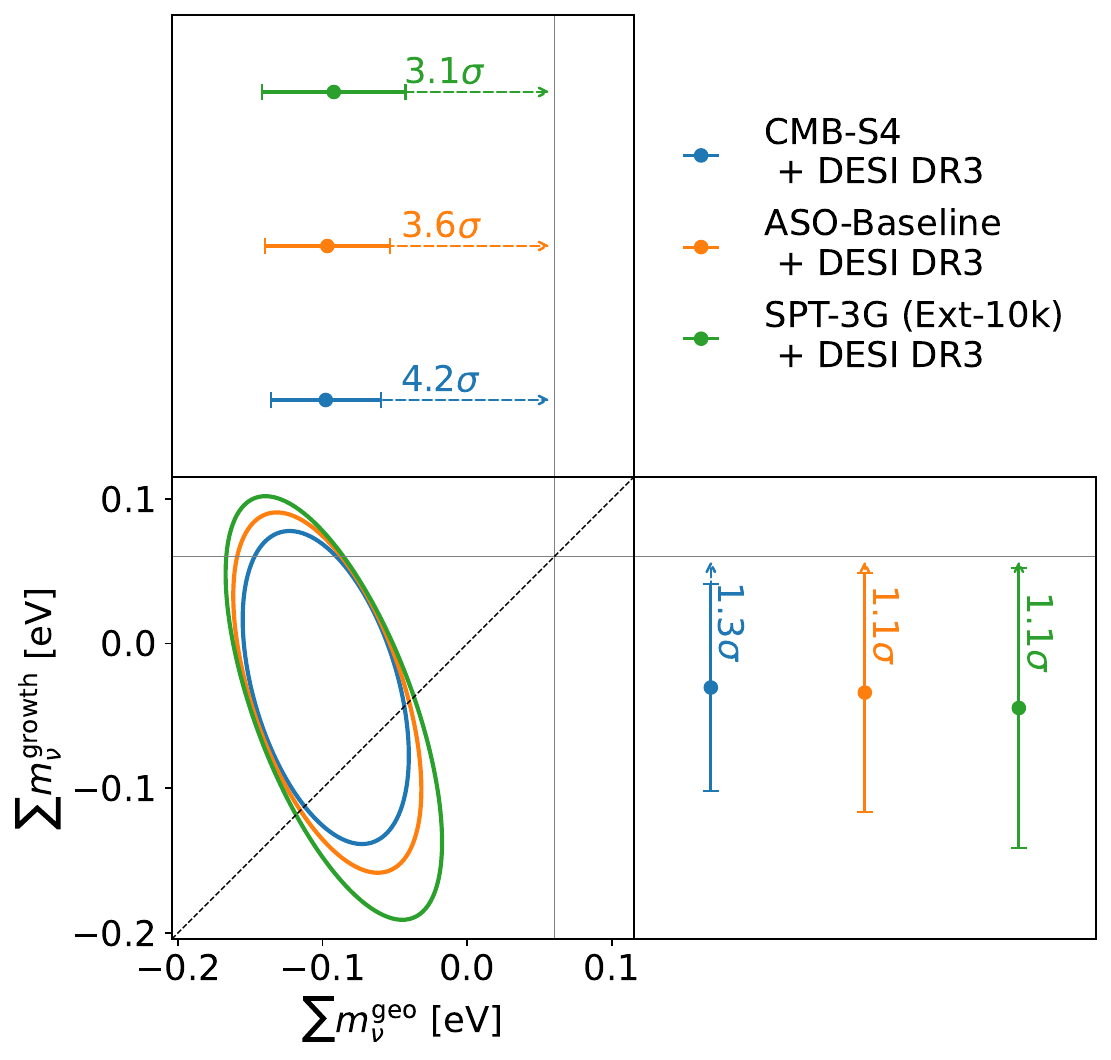}
    \caption{
    Bias from unmodeled evolving dark energy.
    Forecasted bias on $\mgeo$ and $\mgrowth$ if the true underlying cosmology is $w_0w_a$CDM cosmology but the assumed model for inference is $\Lambda$CDM+$\mgeo$+$\mgrowth$. The significance numbers quoted with dashed arrows are the deviation from the fiducial value of 0.058 eV (gray solid lines). The contours in the bottom left panels show the 68\% limit. The dash black line in the bottom left panel shows the constraint of $\mgeo=\mgrowth$.}
    \label{fig:w0wa_bias}
\end{figure}

We next perform a Fisher bias forecast for the case in which the true cosmology is $w_0w_a$CDM with $\mgeo = \mgrowth = 0.058$~eV but the model assumed in the inference is $\Lambda$CDM$+\mgeo+\mgrowth$. For the fiducial $w_0w_a$CDM cosmology we take the DESI DR2 best fit, $w_0 = -0.426$ and $w_a = -1.72$. The resulting shifts are shown in Fig.~\ref{fig:w0wa_bias}. Both parameters are biased low, but very unequally: $\mgeo$ is displaced from the fiducial $0.058$~eV by $3.1\sigma$, $3.6\sigma$ and $4.2\sigma$ for SPT-3G, ASO and CMB-S4 combined with DESI DR3 BAO, while $\mgrowth$ moves by only $1.1$--$1.3\sigma$. This is the expected asymmetry: a wrong dark energy equation of state misrepresents the late-time expansion history, which is exactly what $\mgeo$ absorbs, whereas the perturbation evolution that $\mgrowth$ controls is affected only indirectly, in agreement with \cite{Qu:2026fkt}. 

\begin{figure}
    \centering
    \includegraphics[width=0.95\linewidth]{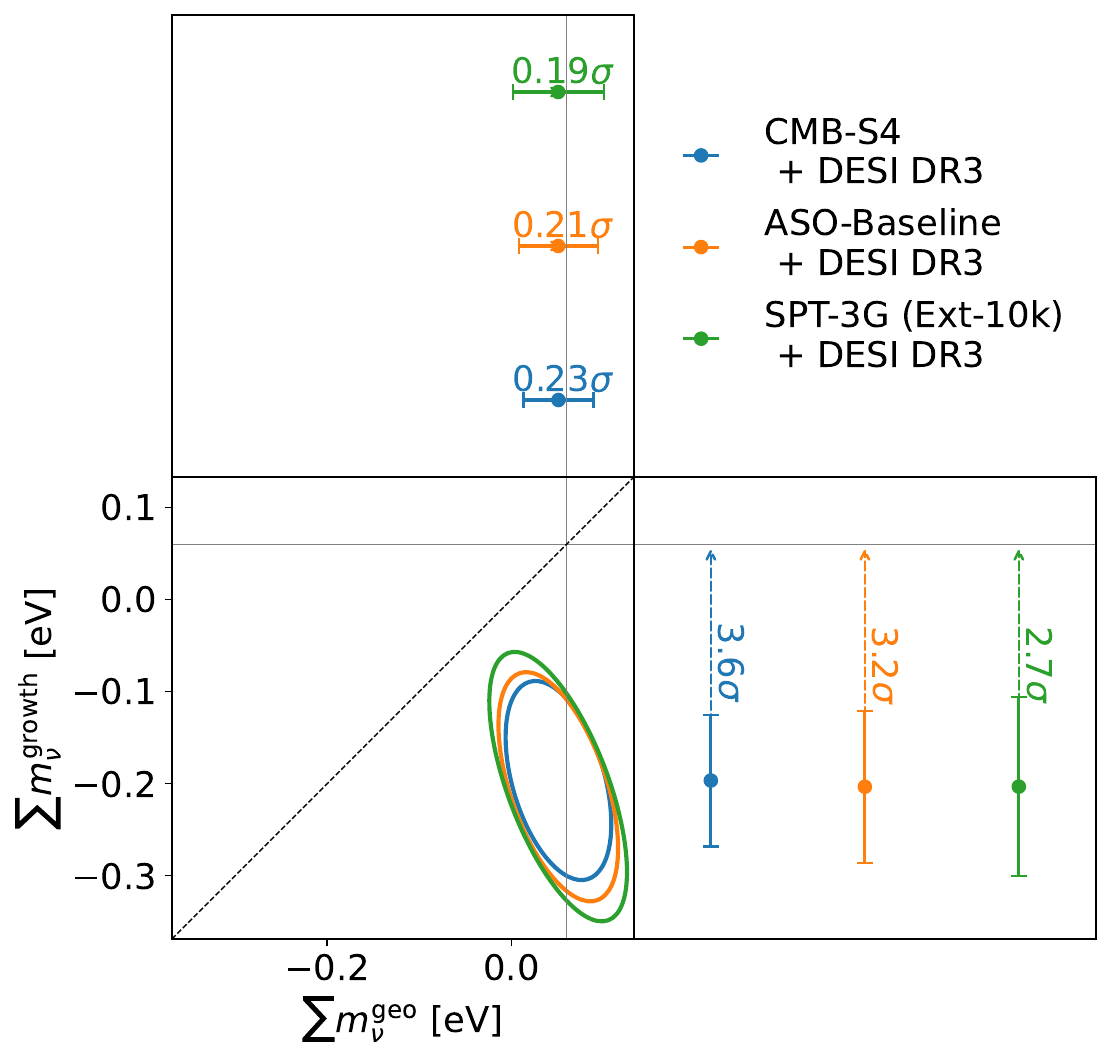}
    \caption{
    Bias from incorrect optical depth.
    Forecasted bias on $\mgeo$ and $\mgrowth$ if the true underlying cosmology is $\Lambda$CDM with $\tau\sim0.09$ but one uses \texttt{SRoll2.EE} low-$\ell$ likelihood enforcing $\tau\sim0.054$. The significance numbers quoted dashed arrows are the deviation from the fiducial value of 0.058 eV, which are shown in Gray solid lines. The contours in the bottom left panels show the 68\% limit. The dash black line in the bottom left panel shows the constraint of $\mgeo=\mgrowth$.}
    \label{fig:tau_bias}
\end{figure}

On the other hand, a complementary case is shown in Fig.~\ref{fig:tau_bias}, where the true cosmology is $\Lambda$CDM with $\tau \sim 0.09$ but the inference uses the \texttt{SRoll2.EE} low-$\ell$ likelihood, which enforces $\tau \sim 0.054$. Here the pattern is reversed: $\mgrowth$ is biased towards lower values from $0.058$~eV by $2.7\sigma$, $3.2\sigma$ and $3.6\sigma$, while $\mgeo$ moves by at most $0.23\sigma$. Again this follows from where the contaminant acts. The optical depth is degenerate with $A_s$ and hence with the amplitude of structure, so an error in $\tau$ propagates into the growth channel through the $\tau - A_s - \omega_c - \sum m_\nu$ degeneracy discussed in Sect.~\ref{sec:param_deg}, while leaving the background expansion untouched.

Taken together, the two cases give the split its diagnostic value. The two leading proposed resolutions of the neutrino mass tension, dynamical dark energy and a higher reionization optical depth, leave opposite and well separated signatures in the pair $(\mgeo, \mgrowth)$. With future data, the pattern of the biases in the $(\mgeo, \mgrowth)$ pair identifies which, if either, is at work. 
A single effective $\meff$ cannot make this distinction, since it responds to both.

It is worth contrasting this diagnostic with the other $\Delta \sum m_{\nu} \equiv \mgeo - \mgrowth$ diagnostic presented earlier.
In both Figs.~\ref{fig:w0wa_bias} and \ref{fig:tau_bias}, the biased contours remain compatible with the line $\mgeo = \mgrowth$: for the $w_0w_a$ case we find $\mdiff = -0.067 \pm 0.095$~eV, a $\sim 0.7\sigma$ departure from zero, and for the $\tau$ case $\mdiff = 0.25 \pm 0.095$~eV departing from 0 at $\sim 2.6 \sigma$ level. The consistency test of Sect.~\ref{sec:result_real} therefore has limited power against some particular contaminants, because the induced shifts to $\mgeo$ and $\mgrowth$ do not depart significantly from the degeneracy direction of the posterior in the $w_0w_a$CDM case. What does discriminate is the offset of each channel from the value expected from oscillation experiments: a large deviation in $\mgeo$ with a small one in $\mgrowth$ points to the background model, and the reverse points to the low-$\ell$ likelihood and $\tau$. Quoting both marginal offsets, rather than only $\mdiff$, is thus essential to using the split as a diagnostic.

\section{Discussion and Conclusion}\label{sec:conclusion}

In this work, we study the origin of constraints on the neutrino mass sum from the CMB and BAO observables, trying to answer which channel, the geometry or growth information, drives the tight constraint on the $\sum m_\nu$ upper bound against the minimum $\sum m_\nu$ from neutrino oscillation experiments. 
We find that both geometry and growth channels have similar constraining power and preference for ``negative'' values of $\sum m_\nu$ with
$\mgeo = -0.07\pm0.10$~eV (68\% C.L.) 
and 
$\mgrowth = -0.13^{+0.10}_{-0.12}$~eV (68\% C.L.). 

The tight upper limit on $\sum m_\nu$, when enforcing the single parameterization of neutrino mass sum without extrapolation to the ``negative'' regime, is not easily attributable to one anomalous dataset, but to the mild $\Omega_m$ disagreement between CMB and BAO in $\Lambda$CDM, in agreement with \cite{Loverde:2024nfi,lynch2025whatsmattersigmamnu}. 
Indeed, in Sect.~\ref{sec:fiducial_swap}, we start with replacing the data vector of each likelihood in turn with mock data free of anomalies or fluctuations. 
When the \texttt{CamSpec} best fit is taken as the mock truth, substituting the BAO data vector returns the $\sum m_\nu$ posterior to a peak near the fiducial $0.058$~eV, with CMB lensing playing a sub-dominant role. 
If instead the joint \texttt{CMB+BAO} best fit is used to generate the mock data, no single substitution (e.g., CMB lensing) changes the bound appreciably. 
Since \texttt{BAO} carries only geometrical information while \texttt{CMB} carries both, splitting $\sum m_\nu$ by dataset cannot answer which channel drives the preferred lower $\sum m_\nu$ constraint, and thus motivates the split by physical effects.

We thus implement a phenomenological split of the neutrino mass sum based on their effects on geometry $\mgeo$ and growth $\mgrowth$ in Sect.~\ref{sec:split_imp}.
$\mgeo$ sets the expansion rate used for geometrical quantities, including the BAO observables, the geometric factors in the CMB source terms and the lensing kernel, while $\mgrowth$ sets the expansion rate entering the perturbation evolution of all species and the metric perturbations. 
Although the BAO observables depend solely on $\mgeo$, the computation of CMB and CMB lensing power spectra is performed through matching the comoving $k$ grid and the conformal-time grids via redshift (Fig.~\ref{fig:flowchart}), a construction that reduces to $\Lambda$CDM$+\sum m_\nu$ when $\mgeo = \mgrowth$.

Our $(\mgeo, \mgrowth)$ implementation is a physically consistent way of dissecting the constraining power from the geometry and structure growth on the neutrino mass sum using CMB and BAO measurements. This differs from previous work. 
The split of \cite{graham2025darkforcesgathering} is by observable, CMB lensing versus BAO, so the geometrical information carried by the CMB is assigned to neither parameter; 
the background-only likelihoods of \cite{Loverde:2024nfi,lynch2025whatsmattersigmamnu} marginalize the perturbative information at the likelihood level rather than isolating it;
and the background-versus-perturbation separation of \cite{Bertolez-Martinez:2024wez} only partitions the neutrino sector internally, in which the impacts of massive neutrinos on the perturbation focuses on the neutrino perturbation itself. 

The constraining power of the two channels is limited by different physics. The primary CMB alone constrains neither parameter well: $\mgeo$ enters only through $D_A(z_*)$, where at fixed $\theta_*$ it is absorbed by $H_0$ and a small shift in $\omega_c$, while $\mgrowth$ is measured through the lensing of the primary spectra themselves. 
Adding CMB lensing improves $\mgrowth$; adding BAO improves $\mgeo$ by a much larger factor, since BAO pins the fraction of matter in the total energy independently of the CMB. 
For the full combination of \texttt{CMB+Lensing+BAO} at current measurement precision, the Fisher forecast gives $\sigma(\mgeo) = 0.074$~eV and $\sigma(\mgrowth) = 0.125$~eV [Eqn.~(\ref{eq:fisher_all})], and the MCMC on mock late-time data agrees at the $10\%$ level [Eqn.~(\ref{eq:mock_all})].

Replacing the mock \texttt{Lensing} and \texttt{BAO} data vectors with the real datasets, the geometry and growth channel gives $\mgeo = -0.07 \pm 0.10$~eV (68 \% C.L.) and $\mgrowth = -0.13^{+0.10}_{-0.12}$~eV (68 \% C.L.) [Eqn.~(\ref{eq:real_all})] respectively. The geometry and growth channels show preference for $\sum m_\nu$ below the minimal value from neutrino oscillation experiment values at approximately $1.3~\sigma$ and $1.8~\sigma$ level respectively.

The difference between the channels, $\mdiff = 0.07 \pm 0.17$~eV [Eqn.~(\ref{eq:delta})], is consistent with zero at $\sim 0.4\sigma$: the background geometry and the perturbation evolution tell the same story at present sensitivity, and the $\Lambda$CDM internal consistency test is passed. 
However, the $\mgeo=\mgrowth=0.058$~eV still lies outside of the 95\% C.L. region of the $\left(\mgeo, \mgrowth\right)$ joint posterior distribution~(Fig.~\ref{fig:mnu_only_geo_growth}).
Under the canonical non-negative prior we find $\mgeo < 0.081$~eV and $\mgrowth < 0.11$~eV (Eqn.~\ref{eq:nonneg_all}), both above $0.058$~eV, whereas the single-parameter analysis of the same data gives $\sum m_\nu < 0.0597$~eV,
which only appears when enforcing $\mgeo=\mgrowth$.

Turning to future data, the two channels separate more cleanly by which experiment drives them. 
Without BAO, $\mgrowth$ improves steadily with CMB sensitivity, from $0.14$~eV for SPT-3G (Ext-10k) to $0.087$~eV for CMB-S4, while $\mgeo$ remains an order of magnitude weaker. 
Adding DESI DR3 BAO inverts the picture: $\sigma(\mgeo)$ collapses to $0.050$--$0.038$~eV across the three configurations, which then differ from one another by only about $32\%$, so the geometry channel becomes essentially BAO-limited while the growth channel continues to be set by the CMB and CMB lensing. 
Future gains in the two channels will therefore come from different experiments. If the negative central value of $\mgeo = -0.07$~eV persists, ASO and CMB-S4 combined with DESI DR3 would reject $0.058$~eV at $3.0\sigma$ and $3.4\sigma$ from the geometry channel alone.

Finally, the split has diagnostic value that a single effective mass does not, because the two leading proposed resolutions of the neutrino mass tension leave opposite signatures in the pair $(\mgeo, \mgrowth)$. 
If the true cosmology is $w_0w_a$CDM preferred by DESI DR2 BAO but $\Lambda$CDM is assumed, $\mgeo$ is biased low by $3.1$--$4.2\sigma$ across our three future configurations while $\mgrowth$ moves by only $1.1$--$1.3\sigma$, because a wrong dark energy equation of state misrepresents exactly the late-time expansion history that $\mgeo$ absorbs. 
If instead the true optical depth is $\tau \sim 0.09$ but the inference enforces $\tau \sim 0.054$ by using the CMB low-$\ell$ likelihood, $\mgrowth$ is biased low by $2.7$--$3.6\sigma$ and $\mgeo$ by at most $0.23\sigma$, since $\tau$ propagates into the amplitude of structure through the $\tau - A_s - \omega_c - \sum m_\nu$ direction while leaving the background expansion untouched. 
With future data the pattern of the two offsets, the geometry and growth channels, therefore identifies which, if either, contaminant is at work.

\section*{Acknowledgement}
The authors acknowledge the High Performance Computing Core Facility at the University of California, Davis, for providing computational resources that have contributed to the research results reported in this paper. 
Computations were also performed on the Niagara supercomputer at the SciNet HPC Consortium. SciNet is funded by Innovation, Science and Economic Development Canada; the Digital Research Alliance of Canada; the Ontario Research Fund: Research Excellence; and the University of Toronto.
WLKW acknowledges support from an Early Career Research Award DE-SC0026376 of the Department of Energy.

We thank Lloyd Knox, Noah Sailer, Santiago Agui, Irene Abril-Cabezas and Uendert Andrade for useful discussions.

\bibliography{apssamp}

\end{document}